\documentclass[fleqn,usenatbib]{mnras}
\usepackage{verbatim}
\usepackage{newtxtext,newtxmath}
\usepackage{threeparttable}

\usepackage[T1]{fontenc}
\usepackage{ae,aecompl}

\usepackage{graphicx}	
\usepackage{amsmath}	
\usepackage{amssymb}	

\newcommand{\codename}{\texttt{juliet}}
\newcommand{\planetnameb}{K2-140b}
\newcommand{\starname}{K2-140}

\title[juliet]{\texttt{juliet}: a versatile modelling tool for transiting and non-transiting exoplanetary systems}

\author[Espinoza et al.]{
N\'estor Espinoza$^{1,2}$\thanks{E-mail: nespinoza@stsci.edu (NE)}\thanks{Bernoulli Fellow}\thanks{IAU-Gruber Fellow},
Diana Kossakowski$^{1}$,
Rafael Brahm$^{3,4,5}$
\\
$^{1}$ Max-Planck-Institut f\"ur Astronomie, K\"onigstuhl 17, 69117 Heidelberg, Germany.\\
$^{2}$ Space Telescope Science Institute, 3700 San Martin Drive, Baltimore, MD 21218, USA.\\
$^{3}$ Center of Astro-Engineering UC, Pontificia Universidad Cat\'olica de Chile, Av. Vicu\~{n}a Mackenna 4860, 7820436 Macul, Santiago, Chile,\\
Av. Vicu\~na Mackenna 4860, 782-0436 Macul, Santiago, Chile.\\
$^{4}$ Instituto de Astrof\'isica, Facultad de F\'isica, Pontificia Universidad Cat\'olica de Chile,\\
Av. Vicu\~na Mackenna 4860, 782-0436 Macul, Santiago, Chile.\\
$^{5}$ Millennium Institute of Astrophysics, Av. Vicu\~na Mackenna 4860, 782-0436 Macul, Santiago, Chile.\\
}

\date{Accepted XXX. Received YYY; in original form ZZZ}

\pubyear{2019}

\begin{document}
\label{firstpage}
\pagerange{\pageref{firstpage}--\pageref{lastpage}}
\maketitle

\begin{abstract}
Here we present \codename, a versatile tool for the analysis of transits, radial-velocities, or both. \codename\ is built over 
many available tools for the modelling of transits, 
radial-velocities and stochastic processes (here modelled as 
Gaussian Processes; GPs) in order to deliver a tool/wrapper which can be used 
for the analysis of transit photometry and radial-velocity measurements from multiple instruments at the same time, using nested 
sampling algorithms which allows it to not only perform a 
thorough sampling of the parameter space, but also to 
perform model comparison via bayesian evidences. In 
addition, \codename\ allows to fit transiting and non-transiting 
multi-planetary systems, and to fit GPs which might share hyperparameters between the photometry and radial-velocities 
simultaneously (e.g., stellar rotation periods), which might be 
useful for disentangling stellar activity in radial-velocity 
measurements. Nested Sampling, Importance Nested Sampling and 
Dynamic Nested Sampling is performed with publicly 
available codes which in turn give \codename\ multi-threading options, allowing it to scale 
the computing time of complicated 
multi-dimensional problems. We make \codename\ publicly 
available via GitHub.
\end{abstract}




\section{Introduction}
The pioneering efforts from both ground-based radial-velocity \citep[see, e.g.,][]{coralie:2000,tinney:2001,harps:2004,sophie:2009,arriagada:2011,butler:2017,CARMENES:2018} and transit 
\citep[see, e.g.][]{pollaco:2006,mearth:2008,ngts:2013,burdanov:2017,pepper:2018,bakos:2018} surveys, along with ambitious 
space-based transit searches \citep{kepler,tess}, has 
transformed and nurtured the known field today
of extrasolar planets. These efforts have not only 
increased the number of known 
planets outside of the solar system from 
just a couple to thousands\footnote{Over 3000 at the time of writing according to \url{http://www.exoplanets.org/} \citep{exorg}.} in about a decade, but have also unveiled 
the diversity of worlds orbiting stars other 
than our Sun.

The data with which the vast majority of the above mentioned discoveries 
has been made is, in general, extensive, and has given rise to various 
analysis tools at different steps of the process. These tools range 
from detection and validation algorithms that help to reject 
possible false-positive scenarios and/or disentangle the planetary 
signals from other possible signals causing the observed data \citep[see, e.g., ][]{Hartman:2009,vespa1,pastis:2014,vespa2,kima:2018} to 
analysis tools used to retrieve the physical and orbital parameters 
of the discovered exoplanets from observed lightcurves \citep[see, e.g.,][]{Gazak:2012,pytransit:2015}, radial-velocities \citep[see, e.g.,][]{systemic,WH:2009,Baluev:2013,IM:2015,malavolta:2016,radvel:2018,kima:2018} or both \citep[see, e.g.,][]{Bakos:2010,Hartman:2012,exonailer,Baluev:2018,pyaneti:2019, allesfitter, exoplanet}, 
some of which even include the modelling of the stellar properties jointly 
with the modelling of the photometry and radial-velocities \citep{eastman:2013,exofast2, exofast:2019,Hartman:2018}.

As can be seen from the above, a plethora of tools are available in 
the literature to perform analyses of exoplanetary signals in order to 
constrain the physical parameters of an orbit given either photometry, 
radial-velocities or both. In terms of planet discovery 
and characterization, tools that can fit both photometry and 
radial velocities coming from different instruments simultaneously will 
be extremely important, as missions like the \textit{Transiting Exoplanet 
Survey Satellite} \citep[TESS;][]{tess} are already starting to 
provide thousands of new interesting exoplanets orbiting bright stars 
which will receive extensive follow-up, especially from 
radial-velocity facilities, that needs to be analyzed in detail in 
order to reveal the physical parameters of the transiting object.

{In this work we introduce an open source library developed to 
perform exactly the kind of analysis discussed above, which has 
many differences and improvements over other available tools 
that perform similar tasks \citep[e.g.,][]{eastman:2013,exofast2,exofast:2019,Baluev:2018,pyaneti:2019}. First, 
this new library allows for the fitting of any number of transiting 
and non-transiting systems, allowing a simultaneous fit of the 
available data coming from either photometry, radial-velocities or 
both. It also allows Gaussian Processes \citep{RW:2006} to be fitted both to the photometry (with 
a different gaussian process to each photometric instrument) and the radial velocities, 
allowing even for common hyperparameters of these Gaussian Processes to be 
shared between both datasets --- useful, e.g., in settings in which rotational modulation information 
(such as a characteristic stellar rotation period) is present both in the photometry and in the 
radial-velocities. In addition, our procedures are efficient at exploring the whole parameter space 
thanks to nested sampling algorithms used in the parameter exploration procedure which, due to 
their nature, allow us to also estimate the probability of different models given the data through 
bayesian evidences (e.g., eccentric orbits, additional planets in the system, different model for 
systematic trends, etc.). This library, \codename, is publicly available at 
GitHub\footnote{\url{https://github.com/nespinoza/juliet}} and is written in Python.} 

{Before detailing the procedures used by \codename\ to model the data and showcase the 
types of analysis it can do, we would like to first motivate how it differentiates itself from 
existing codes that perform similar tasks and, thus, why such a library is needed}. To date and to 
our knowledge, the {seven} open-source tools that can perform both photometric and radial-velocity 
analysis in order to constrain the physical and orbital parameters of a transiting exoplanet are 
\texttt{EXOFAST} \citep{eastman:2013,exofast2, exofast:2019}, 
\texttt{exonailer} \citep{exonailer}, \texttt{PlanetPack} \citep{Baluev:2018}, \texttt{pyaneti} \citep{pyaneti:2019}, PyORBIT \citep{malavolta:2016,malavolta:2018}, {\texttt{allesfitter} \citep{allesfitter} and \texttt{exoplanet} \citep{exoplanet}}. \texttt{EXOFAST} 
is one of the most versatile of the tools described: it indeed allows to fit photometry and radial-velocities 
from different instruments, and even allows to perform the modelling of the stellar properties jointly with 
the available photometry and radial-velocity measurements. However, one of its weaknesses is its inability 
to account for different noise processes in both the photometry and radial-velocities such as Gaussian 
Processes \citep{RW:2006}, which might impact directly on, e.g., the type of radial-velocity signals that it can 
handle and in the type of photometry that is able to simultaneously detrend, which in some 
cases might not be well modelled by the simple linear models in the parameters that it can currently 
handle. \texttt{exonailer} \citep{exonailer}, although allowing to fit radial-velocities and 
photometry from different instruments as well, it is not as versatile as \texttt{EXOFAST}. It does allow to 
detrend photometry in its most recent versions via Gaussian Processes, but because it uses \texttt{emcee} 
\citep{emcee} to perform the parameter exploration, is very sensitive to the initial parameters, often making 
this simultaneous detrending procedure of the photometry unfeasible to be performed jointly with the transit 
and radial-velocity parameter optimization. In addition, \texttt{exonailer} is currently able to fit only one 
planet at the time, and does not allow to model Gaussian Processes in the radial-velocities either. 
{\texttt{PyORBIT} \citep{malavolta:2016,malavolta:2018} does allow to fit transits and radial-velocities 
simultaneously, allowing to fit multiple-planets simultaneously. It does also incorporate Gaussian Processes 
but it also uses \texttt{emcee} for the parameter exploration, which makes it prompt to the same problems of 
initial values as \texttt{exonailer}}. \texttt{PlanetPack} in its most recent version \citep{Baluev:2018} incorporates the modelling of Gaussian Processes 
for the radial-velocity modelling, and allows to fit for a variety of effects including simultaneous transit 
fitting along with the radial-velocity optimization. This code, however, only allows to account for systematic 
trends in the photometry via {polynomials}, and does not allow to fit multiplanetary systems in the 
photometry. The recently introduced \texttt{pyaneti} \citep{pyaneti:2019}, although 
allowing to fit multiple planets in both the radial-velocities and transits, 
it does not allow to handle data obtained from multiple 
photometric instruments, and does not support the use of Gaussian Processes either in the photometry nor in the radial-velocities. {Finally, \texttt{allesfitter} \citep{allesfitter} and \texttt{exoplanet} \citep{exoplanet} are both 
very similar to \texttt{juliet} --- these are projects that were, in fact, 
developed at about the same time than this one. While \texttt{allesfitter} at 
the time of writing is able to fit for a wide range of phenomena 
\textit{including} transit and radial-velocities, we believe 
\texttt{juliet} is much more user-friendly given its thorough documentation\footnote{\url{https://juliet.readthedocs.io/en/latest/}} 
and importability into \texttt{python} scripts. While these latter benefits 
are shared with \texttt{exoplanet}, \texttt{juliet} is an excellent alternative 
if what one is looking for is to perform model comparison via bayesian 
evidences (see below) \textit{and} posterior sampling --- currently, only the 
latter is allowed by \texttt{exoplanet}}. 

There is one additional weakness shared by {almost} all the open source joint 
fitting tools just described above {(with the only exception being \texttt{allesfitter})}: none of them provide tools to perform 
formal model comparison between different models used to fit the data. 
Although \texttt{PlanetPack} and \texttt{EXOFAST} do provide (frequentist) 
statistics on the fits performed to the data, it is not straightforward 
to use those to compare different models, as all frequentist statistical 
measures assume an underlying null hypothesis which does not take into 
account either prior information nor the reality that there are more 
models than the one being tested. This is very important for complicated 
tasks such as calculating the evidence for additional planets in either 
the transits and/or the radial-velocities (especially in the case in which 
extra, non-planetary signals such as stellar activity might make this 
procedure even more difficult), or even for more simple but 
routine tasks such as finding evidence for an eccentric orbit 
\citep[or even disentangling between significant eccentricity or additional planets; 
see e.g., ][]{Kurster:2015}. Although this question has been explored in the literature 
for the detection of planets in radial-velocities \citep[see, e.g. the excellent discussion 
on this topic by][and references therein]{Nelson:2018} and even open 
source tools have been developed to aid in quantifying this evidence 
\citep{kima:2018}, no general open source tool is available for this that 
allows to incorporate transits and radial-velocities coming from data 
from different instruments. {From our discussion, we believe there is thus a strong need for 
a tool that can incorporate both linear and Gaussian Process regression in both transits and radial-velocities, 
that can take into account multiple-planet systems, and that can, on top of that, provide a quantitative 
measure of the evidence of different models so that they can be compared and be either selected or 
combined in a formal manner. This is the main motivation behind \codename.}

{This} work, {which introduces our \codename\ library,} is organized as follows. 
Section \ref{sec:modelling} presents how the data modelling is treated within \codename. In 
Section \ref{sec:tests} we present some tests of the code 
on real data, with which we show some its capabilities. Section 
\ref{sec:dc} presents a discussion and in \ref{sec:cfw} we present our 
conclusions and future work.

\section{Data modelling within \texttt{juliet}}
\label{sec:modelling}

{In this section, we introduce the probabilistic models \codename\ assumes when performing photometric and/or radial-velocity fits to data. For both of those types 
of datasets, \codename\ considers a common model in 
which each datapoint $y(t_{i,l})$ at time $t_{i,l}$, 
with $i$ being and index that identifies each instrument 
and $l$ being an index identifying a datapoint in a 
given instrument (i.e., $l\in [0,1,...,N_i]$, where $N_i$ 
is the total number of datapoints in instrument $i$) is 
given by a probabilistic model of the form:
\begin{equation}
\label{eq:gm}
y(t_{i,l}) \sim \mathcal{M}_i(t_{i,l}) + \textnormal{LM}_{i}(t_{i,l}) + \epsilon_{i}(t_{i,l}).
\end{equation}
Here, $\mathcal{M}_i(t_{i,l})$ denotes the particular photometric (described in Section \ref{sec:photmod}) or radial-velocity (described in Section \ref{sec:rvmod}) 
model for instrument $i$, which 
depends on the physical planetary parameters of the 
system being modelled (e.g., the planet-to-star radius ratio 
for transits or the semi-amplitude for radial-velocities) as well as instrumental parameters (e.g., the limb-darkening parameters for transit models, or the systemic velocities of 
each instrument for radial-velocities). 
$\textnormal{LM}_{i}$ is a linear model for instrument 
$i$ of the form
\begin{eqnarray}
\label{eq:lm}
\textnormal{LM}_{i}(t_{i,l}) = \sum^{p_i}_{n=0} x_{n,i}(t_{i,l})\theta^{\textnormal{LM}}_{n,i},
\end{eqnarray}
where the $x_{n,i}(t_{i,l})$ are the $p_i+1$ linear 
regressors at time $t_{i,l}$ for instrument $i$, and 
$\theta^{\textnormal{LM}}_{n,i}$ are the coefficients of 
those regressors (e.g., $x_{n,i}(t_{i,l}) = t_{i,l}^n$ would 
model a polynomial trend for instrument $i$). Finally, $\epsilon_{i}(t_{i,l})$ is a zero-mean noise term, 
which \codename\ can model in various forms including 
Gaussian Processes \citep[GPs;][see Section \ref{sec:GP} 
for details on the kernels that \codename\ can handle]{RW:2006}. As will be detailed in Section 
\ref{sec:GP}, this term can be either an individual term 
for each instrument, or be common to all instruments, 
where the dependance with each instrument in this latter 
case is only through a jitter term unique to each 
instrument, $\sigma_{w,i}$.}

{Given thus the vector of the \textit{physical parameters of the planets} that define the photometric 
or radial-velocity models (i.e., the vector containing 
all the physical elements that define each model), $\vec{\theta}_\mathcal{P}$, \codename\ can handle two types of 
possible models for either the photometry or the radial-velocities. 
The first is an ``instrument-by-instrument" model in which the log-likelihood of each instrument is 
assumed to be different (i.e., each instrument has its own 
individual noise model with possible common hyperparameters 
with other instruments). In this case, the full log-likelihood 
considering all the instruments is easily separable as a sum of log-likelihoods for each instrument (because they are independant from each other).  If we consider the vector that defines each 
instrumental model $\vec{\theta}_{i}$ (which includes, 
e.g., the coefficients $\theta_{n,i}^{\textnormal{LM}}$ 
of the linear model for each instrument and the hyperparameters of 
the chosen noise model for instrument $i$) and the vector  
$\vec{y}_i = (y(t_{i,0}), y(t_{i,1}),...,y(t_{i,N_i}))^T$ which has the 
probabilistic model for all the datapoints in instrument $i$, then the total log-likelihood of the model considering 
the data of all instruments, $\mathcal{D}_I$, for the whole photometric or radial-velocity model has a common form, given by
\begin{eqnarray}
\label{eq:like}
\ln p(\mathcal{D}_I|\vec{\theta}_\mathcal{P}, 
\vec{\theta}_{0},\vec{\theta}_{1},...) = \sum_{i=0}\ln p(\vec{y}_i|\vec{\theta}_\mathcal{P}, 
\vec{\theta}_{i}),
\end{eqnarray}
where in general we assume the likelihood for each instrument follows 
the likelihood of a $N_i$-dimensional multivariate gaussian, i.e.,  
\begin{eqnarray*}
\ln p(\vec{y}_i|\vec{\theta}_\mathcal{P}, 
\vec{\theta}_{i}) = -\frac{1}{2}\left[N_i\ln 2\pi + \ln\left|\mathbf{\Sigma}_i\right|  + \vec{r}_i^T \mathbf{\Sigma}_i^{-1}\vec{r}_i \right],
\end{eqnarray*}
where each element of vector $\vec{r}_i$, $r_{i,l}$, is 
given by
\begin{eqnarray*}
r_{i,l} = y_i(t_{l,i}) - \mathcal{M}_i(t_{i,l}) - \textnormal{LM}_{i}(t_{l,i}).
\end{eqnarray*}
The second type of model \codename\ is able to handle is a so-called 
``global" model, in which the noise model for either the whole 
photometric or radial-velocity dataset is \textit{common} to all 
instruments. In this case, $\epsilon_i(t_{i,l}) \equiv \epsilon(t_{i,l}) + \bar{\epsilon}_i(t_{i,l})$, where $\bar{\epsilon}_i(t_{i,l})\sim N(0,\sigma^2_{w,i}+\sigma^2_{t_{i,l}})$, and where $N(\mu,\sigma^2)$ denotes a normal distribution with mean $\mu$ 
and variance $\sigma^2$. Here, $\sigma^2_{t_{i,l}}$ are the formal 
uncertainties for datapoint $y(t_{i,l})$, and $\sigma^2_{w,i}$ is a 
jitter term that can be defined or fitted for instrument $i$.}

{Here, the whole dataset $\mathcal{D}_I$ is 
modelled together without specific noise models on each 
instrument (appart for the white-noise part individual to 
each instrument, $\bar{\epsilon}_i(t_{i,l})$). Considering 
the stacked data vector $\vec{y}$ containing all the data 
$y(t_{l,i})$ for all instruments, and the stacked parameter 
vector containing all the physical and instrumental parameters, $\vec{\theta}$, the 
total log-likelihood is in this case of the form
\begin{equation}
\label{eq:like2}
\ln p(\vec{y}| \vec{\theta}) = -\frac{1}{2}\left[N_I\ln 2\pi + \ln\left|\mathbf{\Sigma}_I\right|  + \vec{r}^T \mathbf{\Sigma}_I^{-1}\vec{r} \right],
\end{equation}
where $N_I = \sum N_i$. Here, each element of the residual vector 
$\vec{r}$, $r(t_{i,l})$ is given by
\begin{eqnarray*}
r(t_{i,l}) = y(t_{i,l}) - \mathcal{M}_i(t_{i,l}) - \textnormal{LM}_{i}(t_{i,l}).
\end{eqnarray*}
In this case, thus, the elements of the 
covariance matrix $\mathbf{\Sigma}_I$ are given by
\begin{eqnarray*}
\mathbf{\Sigma}_I(t_{i,l},t_{j,m}) = k(x_{i,l},x_{j,m}) + (\sigma^2_{w,i} + \sigma^2_{t_{i,l}})\delta_{t_{i,l},t_{j,m}} 
\end{eqnarray*}
with $\delta_{t_{i,l},t_{j,m}}$ a Kronecker's delta, and $k(\cdot)$ 
being either zero for a pure white-noise model, or equal to any of 
the kernels defined in Section \ref{sec:GP}.}

{It is important to distinguish the physical significance 
of the ``instrument-by-instrument" model given in equation (\ref{eq:like}) and the ``global" model given in equation (\ref{eq:like2}). Because the former assumes a different noise model 
for each instrument, this model assumes each instrument provides 
a distinct realization of a noise process. That is, even if two 
instruments share \textit{all} the hyperparameters of the noise 
model, those two instruments are modelled as if their observed 
noise processes were generated by different realizations of the same process. The ``global" model, however, assumes that not only 
all instruments share the hyperparameters of the selected noise 
model: it assumes \textit{all come from exactly the same realization 
of the process}. This latter model, thus, is typically very useful 
for observations of physical processes (e.g., rotational modulation 
in either photometry or radial-velocities) with the same instrument 
over different seasons, and/or similar instruments, whereas the 
former model is typically useful for observations of instruments 
with different bandpasses and/or different underlying noise 
processes.}

\subsection{Photometric modelling}
\label{sec:photmod}
The {model within} \codename\ 
{assumed for the photometry for each instrument, $\mathcal{M}_i(t_{i,l})$, is of the form}
\begin{eqnarray}
\label{eq:photmodel}
\mathcal{M}_i(t_{i,l}) = \left[\mathcal{T}_i(t_{i,l}) D_i + (1-D_i)\right]\left(\frac{1}{1+D_iM_i}\right).
\end{eqnarray}
{Here,} $\mathcal{T}_i(t_{i,l})$ is the full transit 
model including any number of $N^\mathcal{T}_p$ 
planets in the system for instrument $i$, $D_i$ is a 
dilution factor for the given instrument and $M_i$ is a mean offset out-of-transit flux. {The parameters that define 
the photometric model (described in detail below) are given in Table \ref{tab:parameters-photometry}.}

\begin{table}
    \centering
    \caption{List of parameters that define the model $\mathcal{M}_i(t_{i,l})$ for the photometry for instrument $i$. Note that the same planetary parameters have to be given to every planet in the system.}  
    \label{tab:parameters-photometry}
    \begin{tabular}{lcl} 
        \hline
        \hline
        Parameter name & Units & Description \\
                \hline
                \hline
        Planetary parameters & & \\
        ~~~$P$ & days & Orbital period. \\
        ~~~$t_{0}$ & days & Time of transit-center. \\
        ~~~$p$ & --- & Planet-to-star radius ratio$^1$. \\
        ~~~$b$ & --- & Impact parameter$^1$. \\
        ~~~$a/R_*$ & --- & Scaled semi-major axis$^2$. \\
        ~~~$e$ & --- & Eccentricity of the orbit$^3$. \\
        ~~~$\omega$ & deg & Argument of periastron$^3$. \\
        Instrumental parameters$^4$ & & \\
        ~~~$D_i$ & --- & Dilution factor. \\
        ~~~$M_{i}$ & relative flux & Relative flux offset. \\
        ~~~$q_{1,i}$ & --- & Limb-darkening parameter$^5$. \\
        ~~~$q_{2,i}$ & --- & Limb-darkening parameter$^5$. \\
        \hline
        \hline
    \end{tabular}
    \begin{tablenotes}
      \small
      \item $^1$ Instead of fitting directly for $(p,b)$, \codename\ allows to fit for $(r_1,r_2)$, which samples 
      the whole range of physically plausible values for $p$ and $b$ \citep[see][for details]{Espinoza:2018}.
      \item $^2$ Instead of the scaled semi-major axis, the stellar density $\rho_*$ (in units of kg/m$^3$) can be fitted. If this is the case, for multiple-planet fits only one value is needed to constrain $a_k/R_*$ of all the 
      $k$ planets (see text).
      \item $^3$ Instead of fitting for $(e,\omega)$ directly, \codename\ allows to fit for the first and second 
      Laplace parameters $\mathcal{E}_1 = e\sin \omega$ and $\mathcal{E}_2 = e\cos \omega$ or the parameters $\mathcal{S}_1=\sqrt{e}\sin \omega$ and $\mathcal{S}_2=\sqrt{e}\cos \omega$. This latter parametrization is recommended.
      \item $^4$ In addition, the coefficients for the linear 
      models, $\theta^{\textnormal{LM}}_{n,i}$, can optionally be defined if linear regressors are fed to \codename.
      \item $^5$ Note these are \textit{not} the limb-darkening coefficients, except for the linear law, where $q_2 = 0$ and $q_1 = u$, the linear limb-darkening coefficients. The mapping of these parameters to the limb-darkening coefficients is described in \cite{kipping:2013} and \cite{EJ:2016} (see text).
    \end{tablenotes}
\end{table}

{The motivation behind the model defined in (\ref{eq:photmodel})} is that 
precise photometric instruments like \textit{TESS} will provide 
photometry that will be contaminated by the flux of nearby sources 
due to its large pixel size \citep[e.g., 21'' for TESS,][]{tess}, which will in 
turn contaminate the ``true" transit parameters due to this dilution 
of the transit shape by nearby light sources to the target star. 
In addition, the model is also motivated by the work 
of \cite{KT:2010} which predicts 
so-called ``self-dilution" of a planet might happen due to light 
from the exoplanet's night-side diluting the transit signature. To understand the process 
of this dilution, how this impacts on the transit model $\mathcal{T}(t)$ and what the 
parameters in equation (\ref{eq:photmodel}) mean, let $F_T$ be the out-of-transit flux of the 
target star in a given passband and let $\sum_{n} F_n$ be the total 
flux of any $n$ other sources in the photometric aperture used to obtain 
the observed flux of the target as a function of time, $F_O (t)$ (note 
that among the $n$ sources in this formalism, a subset might actually 
be the flux of the planets transiting the target star themselves, 
which would give rise to the ``self-dilution" effect discussed 
in \cite{KT:2010}). This latter flux will thus be given by
\begin{eqnarray*}
F_O(t) =  \mathcal{T}(t)F_T + \sum_{n} F_n.
\end{eqnarray*}
In this formalism, $\mathcal{T}(t)$ will be one for out-of-transit 
times $t$ and less than one at in-transit times, implying that 
the \textit{physical} out-of-transit flux will be simply 
$F_T + \sum_{n} F_n$. Typically, if available, one would estimate 
this out-of-transit flux (via, e.g., the mean, median, etc. of 
the out-of-transit measurements) and 
compute relative fluxes by simply dividing this estimate to the 
observed flux $F_O(t)$. However, in reality there is no guarantee 
there will actually be out of transit flux in order to estimate 
it from the data (from, e.g., follow-up transit lightcurves), and 
thus one performs an estimate of the out-of-transit flux on the 
data that might deviate from this physical picture. Let us assume 
the \textit{estimated} out-of-transit flux is of the form 
$F_T + \sum_{n} F_n + E$, where $E$ is a real constant representing 
the offset flux from the real out-of-transit flux (note this flux 
could be negative; e.g., in the case in which this estimate comes 
from the median of the observations of a transit with no or very 
little out-of-transit measurements, or where it is not clear where 
the out-of-transit flux actually is located in time). With this, 
the relative flux $\hat{F}_O(t) = F_O(t)/(F_T + \sum_{n} F_n + E)$ 
will thus be given (after some rearrangements) by
\begin{eqnarray}
\label{eq:photmotivation}
\hat{F}_O(t) = \left[\mathcal{T}(t)D + (1-D)\right]\left(\frac{1}{1+D(E/F_T)}\right),
\end{eqnarray}
where
\begin{eqnarray*}
D = \frac{1}{1 + \sum_n F_n/F_T}.
\end{eqnarray*}
Comparing equation (\ref{eq:photmotivation}) with {equation} (\ref{eq:photmodel}), one can see that if the photometric datapoints $y_(t_{i,l})$ are relative 
fluxes, {in the absence of a linear model} one 
can physically interpret the expected values of $D_i$ and $M_i$ in equation (\ref{eq:photmodel}) directly with the terms in (\ref{eq:photmotivation}). On one hand, because $\sum_n F_n/F_T > 0$, 
this implies that the dilution factor $0 \leq D_i \leq 1$ and {thus} that \textit{the smaller $D_i$ is, the 
largest the dilution by instrument $i$ is}. This parametrization for the 
dilution factor $D_i$ is thus very useful because it gives a strict 
support to this parameter. As for $M_i$, this factor controls and 
fits for the relative (and arbitrary) offset (with respect to the 
target flux) that was applied to generate the observed data 
$y_i^{P}(t_l)$. This offset (relative, with respect to the target) 
flux can vary in sign and value, 
but is always around zero for data whose relative fluxes have 
been estimated with plenty of out-of-transit flux; in the worst 
case scenarios, this factor can be of the same of order as the 
transit depth. In general, $D_iM_i>-1$.
It is interesting 
to note that given this physical interpretation for $D_i$ and $M_i$, 
one might put priors on these parameters given known apertures and 
diluting sources \citep[e.g., for the case of TESS, using the TESS input catalog;][]{TIC}. However, it is important to note that the diluting sources could also be of instrumental (e.g., miscalculated 
background flux) and/or astrophysical (e.g., as is the 
case for ``self"-dilutions discussed above) origin, and as such care 
must be taken when imposing strong priors on those parameters. 
Similarly, care must be taken if one chooses to put too wide of 
a prior (e.g., uniform between 0 and 1) on this parameter as it 
could lead to, e.g., a large transit depth with a very small 
dilution factor in low signal-to-noise transits. These cases 
should, thus, be always inspected and interpreted in order to 
decide the best model for a given dataset.

{Within \codename}, the transit model $\mathcal{T}_i(t_{i,l})$, is generated by 
using \texttt{batman} \citep{batman:2015}, which has 
many flexible options useful for transit modelling 
including the supersampling of the lightcurve model 
in cases of long-cadence integrations 
\citep[for details, see][]{kipping:2010}, a mode that 
is, thanks to \texttt{batman}, also implemented within \codename. In our 
{library}, the full transit model $\mathcal{T}_i(t_l)$ is actually {obtained by 
subtracting 1 to the transit model of each of the $N^\mathcal{T}_p$ planets, which gives us 
the percentage of light each planet is obscuring; then, all the contributions are added up, 
which gives us the total percentage of light occulted by all the planets. We add 1 to this 
result in order to have a normalized total transit lightcurve. This allows} \codename\ to 
efficiently fit multiplanetary transiting systems {with the caveat that although our code is able to model 
multiple planets obscuring the stellar surface simulatenously, \textit{it is not able to model planet-planet transits \citep[see, e.g., ][and references therein]{planetplanet}}}. The transit parameters that are allowed to vary for each transiting planet $k \in [1,...,N^\mathcal{T}_p]$ (where for simplicity, we 
drop the subscript $k$ to identify its parameters in what follows unless otherwise stated) are the planet-to-star radius ratio, $p=R_p/R_s$, the semi-major axis in stellar units, $a/R_*$, the 
impact parameter of the orbit, $b = (a/R_*)\cos i_p$, where $i_p$ is the inclination of 
the planetary orbit with respect to the plane of the sky, the period of the orbit $P$, the time of transit center $t_0$, the argument of periastron 
passage $\omega$ and the eccentricity of the orbit, $e$. In practice, 
\codename\ allows to parametrize either $p$ and $b$ directly, 
or using the efficient sampling scheme detailed in \cite{Espinoza:2018} in which two parameters $r_1$ and 
$r_2$ defined between 0 and 1 are sampled and which explore all the physically meaningful ranges for $p$ and 
$b$ in the $(b,p)$ plane, which ensures the condition $b < 1 + p$ is always satisfied without the need to 
perform rejection sampling (i.e., reject samples for which $b \geq 1+p$). In addition, and 
for reasons that will be detailed in Section \ref{sec:sdmod}, within \codename\ we also 
allow to fit for a common stellar density 
for all the transiting exoplanets in the 
system, 
$\rho_*$, instead of individual values 
of $a/R_*$ for each transiting exoplanet 
in the system. The eccentricity and 
argument of periastron passage can either be parametrized within \codename\ (1) directly, (2) using the first and second Laplace parameters 
$\mathcal{E}_1 = e\sin \omega$ and $\mathcal{E}_2 = e\cos \omega$, in which case the eccentricity and argument 
of periastron are defined as $e = \sqrt{\mathcal{E}_1^2 + \mathcal{E}^2_2}$ and $\omega = \textnormal{atan2}(\mathcal{E}_1,\mathcal{E}_2)$ \cite[which, as noted by][puts a pathological prior of $p(e)=e$ on the eccentricity]{Ford:2006} or (3) via the transformations $\mathcal{S}_1=\sqrt{e}\sin \omega$ and $\mathcal{S}_2=\sqrt{e}\cos \omega$, in 
which case $e = \mathcal{S}_1^2 + \mathcal{S}^2_2$ and $\omega = \textnormal{atan2}(\mathcal{S}_1,\mathcal{S}_2)$; for an excellent discussion on the advantages and disadvantages of those parametrizations, see \cite{eastman:2013}. It is important to note here that in practice in order to generate a transit model, \texttt{batman} needs the inclination and not the 
impact parameter of a given planet in order to generate its transit model. To 
transform the impact parameter in order to obtain the inclination of the orbit in 
the general case of eccentric orbits, \codename\ calculates the inclination 
as \citep[see, e.g.,][]{winn:2010}:
\begin{eqnarray*}
i_p = \arccos \left[\frac{b}{a/R_*} \left( \frac{1 + e \sin \omega}{1 - e^2}\right)\right],
\end{eqnarray*}
which is, of course, valid as long the term inside the $\arccos$ is $\leq 1$. 
For limb-darkening, \codename\ does not use a 
direct parametrization of the limb-darkening 
coefficients as the philosophy within \codename\ is to, whenever possible, \textit{fit for} the limb-darkening coefficients in the analysis, as the procedure of fixing limb-darkening coefficients to 
values obtained from stellar models is known to give rise to biases in the derived 
transit parameters for precise transit lightcurves such as the ones obtained by the \textit{Kepler} mission and the ones currently being obtained by \textit{TESS} \citep{EJ:2015,EJ:2016}. Moreover, given the 
importance of using different limb-darkening laws for different systems/instruments, \codename\ allows to 
fit any one or two-parameter law available via \texttt{batman} which includes the 
linear, quadratic, square-root and logarithmic laws --- the exponential limb-darkening law is not included in this list as it might give rise to 
unphysical results \citep{EJ:2016}. In practice, for the two-parameter laws, 
\codename\ uses the parametrization 
proposed by \cite{kipping:2013}, where two parameters $q_{i,1}$ and $q_{i,2}$ defined between 0 and 1 are 
sampled (one pair for each photometric instrument $i$ used in the analysis) in order to produce, for a 
given limb-darkening law, only physically plausible limb-darkening coefficients for the selected laws, 
which imply that the intensities are everywhere positive and produce decreasing gradients towards the 
limb of the star. For all the two-parameter laws but the logarithmic \codename\ uses the transformations 
derived in \cite{kipping:2013} to go from the $(q_{i,1},q_{i,2})$ plane to the limb-darkening 
coefficients $(u_{i,1},u_{i,2})$ plane; for the logarithmic law, \codename\ uses the transformations derived in 
\cite{EJ:2016}. For the linear law, \codename\ uses one parameter $q_{i} \equiv u_i$ which defines the 
intensity profile of the star. 
It is important to mention here that while the parameters of a given 
planet in the transit model (i.e., $p$, $a/R_*$, $b$, $P$, $t_0$, $\omega$, $e$) are all the same 
along instruments, \codename\ uses one set of limb-darkening coefficients which is unique to 
each instrument and \textit{common among different planets observed with the same instrument}. This might 
seem intuitive as all the planets in a given system are transiting the same star. However, retrieved 
limb-darkening coefficients are also known to depend on the geometry of the system even \textit{for the 
same star} \citep{Howarth:2011}. In practice, however, we believe the 
errors on the retrieved limb-darkening coefficients are not precise enough to see this latter differences on multiplanetary systems and, as such, decide 
within \codename\ to have one common set of limb-darkening 
coefficients per instrument. \codename\ also allows for each instrument to have its own limb-darkening law, as 
the bias/variance tradeoff on the retrieval of transit parameters from transit lightcurves when fitting 
for the limb-darkening coefficients is known to scale differently depending on the number of datapoints, 
geometry of the system and response function \citep[see, e.g., ][]{EJ:2016}.

\subsection{Radial-velocity modelling}
\label{sec:rvmod}
The {corresponding model $\mathcal{M}_i(t_{i,l})$ used within \codename\ to model the radial-velocities for each instrument is given by}
\begin{eqnarray*}
\mathcal{M}_i(t_{i,l}) = \mathcal{K}(t_{i,l}) + \mu_i + Q\left(t'_{i,l}\right)^2 + At'_{i,l} + B.
\end{eqnarray*}
{Here,} $\mathcal{K}(t_{i,l})$ is a full Keplerian signal including any number of $N^\textnormal{RV}_p$ planets in the system, $\mu_i$ is an instrument-dependant systemic velocity {and} $Q$, $A$ and $B$ define 
optional quadratic and linear terms along with the corresponding intercept, respectively, of a 
long-term trend present on the data (coming from, e.g., additional long-period 
companions/activity whose period is unconstrained by the current data). {This latter trend --- common to all 
instruments --- in turn, 
depends on $t'_{i,l} = t_{i,l} - t_a$, where $t_a$ is 
an arbitrary user-defined time (default within 
\codename\ is $t_a = 2458460$). Table \ref{tab:parameters-rvs} 
lists all the parameters (described in detail below) needed to define the radial-velocity model.}.

\begin{table}
    \centering
    \caption{List of parameters that define the model $\mathcal{M}_i(t_{i,l})$ for the radial-velocities (RVs) for instrument $i$. Note that the same planetary parameters have to be given to every planet in the system.}  
    \label{tab:parameters-rvs}
    \begin{tabular}{lcl} 
        \hline
        \hline
        Parameter name & Units & Description \\
                \hline
                \hline
        Planetary parameters & & \\
        ~~~$P$ & days & Orbital period. \\
        ~~~$t_{0}$ & days & Time of transit-center. \\
        ~~~$K$ & m/s or km/s & RV semi-amplitude. \\
        ~~~$e$ & --- & Eccentricity of the orbit$^1$. \\
        ~~~$\omega$ & deg & Argument of periastron$^1$. \\
        Instrumental parameters$^{2,3}$ & & \\
        ~~~$\mu_i$ & m/s or km/s & Systemic RV. \\
        \hline
        \hline
    \end{tabular}
    \begin{tablenotes}
      \small
      \item $^1$ Instead of fitting for $(e,\omega)$ directly, \codename\ allows to fit for the first and second 
      Laplace parameters $\mathcal{E}_1 = e\sin \omega$ and $\mathcal{E}_2 = e\cos \omega$ or the parameters $\mathcal{S}_1=\sqrt{e}\sin \omega$ and $\mathcal{S}_2=\sqrt{e}\cos \omega$. This latter parametrization is recommended.
      \item $^2$ In addition, the coefficients for the linear 
      models, $\theta^{\textnormal{LM}}_{n,i}$, can optionally be defined if linear regressors are fed to \codename.
      \item $^3$ Additionally, one can fit quadratic (through the parameter $Q$, with units of (m/s)/day$^2$ or (km/s)/day$^2$) and linear trends (through the parameter $A$, with units of (m/s)/day or (km/s)/day$^2$), which in turn can involve an intercept term ($B$, with units m/s or km/s).
    \end{tablenotes}
\end{table}

To compute a model for the Keplerian signal, $\mathcal{K}(t_{i,l})$, \codename\ uses \texttt{radvel} \citep{radvel:2018}, which easily implements any number 
of radial-velocity planetary signals in its modelling, and thus allows us to consider 
multiplanetary signals \citep[as a sum of Keplerians, i.e., neglecting dynamical interactions 
\textit{between} exoplanets; see ][]{laughlin:2003}. 
The parameters that \codename\ uses to define the model for 
the Keplerian of planet $k \in [1,...,N^\textnormal{RV}_p]$ are the semi-amplitude of the 
variation, $K_k$, the period of the orbit, $P_k$, the 
time of transit center, $t_{0,k}$, the argument of 
periastron passage, $\omega_k$ and the eccentricity of 
the orbit, $e_k$ (with these latter two parameters being able to be parametrized 
as defined in the previous sub-section for the photometry). Note that all the 
parameters but $K_k$ are common with the transit model for each planet 
present in the system. Also note that \codename\ allows either $N^\textnormal{RV}_p = N^\mathcal{T}_p$ (i.e., 
both planets transit \textit{and} show radial velocity signatures), 
$N^\textnormal{RV}_p > N^\mathcal{T}_p$ (i.e., 
only a subset of the planets in the system transit) and 
$N^\textnormal{RV}_p < N^\mathcal{T}_p$ (i.e., 
only a subset of the transiting planets show radial-velocity 
signatures).

\subsection{Noise models supported within \codename}
\label{sec:GP}
{As described at the beggining of this Section, currently 
\codename\ allows to adopt different forms for the 
noise model term $\epsilon_{i}(t_{i,l})$ defined in 
equation (\ref{eq:gm}), in order to allow 
flexibility in the modelling structure. The simplest form of this noise 
model is that of a white-noise model. In this case, 
this term is assumed to be of the form}
\begin{eqnarray*}
\epsilon_{i}(t_{i,l})\sim N(0,\sigma_{w,i}^2 + \sigma_{t_{i,l}}^2).
\end{eqnarray*}
Here, $\sigma_{w,i}$ is a {jitter term added in quadrature 
to each of the errorbars of each datapoint, 
$\sigma_{t_{i,l}}$}, which can be left as a free 
parameter {or fixed} in the fit. {This 
very simple form in turn implies that the covariance matrix 
$\mathbf{\Sigma}_i$, needed to evaluate 
equation (\ref{eq:like}), is simply a diagonal matrix 
with terms $\sigma^2_{w,i} + \sigma^2_{t_{i,l}}$ in the 
diagonal, which implies this is the fastest noise 
model \codename\ can currently handle, as it requires 
no matrix inversions.}

{The simple white-noise model, however, is 
inadequate in most cases. Correlated noise-structures, 
quasi-periodic signals and/or systematic trends which 
can only be fit by large-degree polynomials are common 
to find in general. Because of these possibilities, 
within \codename\ one can model the noise term as a GP 
in order to have a non-parametric approach to use 
in those different situations.} If assumed as a multi-dimensional GP, then \codename\ 
assumes $\epsilon_{i}(t_{i,l})\sim \mathcal{GP}(0,\mathbf{\Sigma}_i(\mathbf{X}_i))$, 
where $\mathcal{GP}(\vec{0},\mathbf{\Sigma}_i(\mathbf{X}_i))$ is a multi-dimensional GP with $\mathbf{X}_i$ 
being a $D_i \times N_i$ matrix containing the $D_i$ external parameters that define the GP kernel for 
instrument $i$, which in turn defines the $N_i \times N_i$ covariance matrix of the process 
$\mathbf{\Sigma}_i$. The elements of the covariance matrix (for which we'll drop the $i$ subscript 
for simplicity in what follows, along with the one for the $\mathbf{X}_i$ matrix), are in turn defined to be of the form
\begin{eqnarray}
\label{eq:cov}
\mathbf{\Sigma}_{l,m} = k_i(\vec{x}_l,\vec{x}_m) + (\sigma_{w,i}^2 + \sigma_{t_{i,l}}^2)\delta_{l,m}, 
\end{eqnarray}
where the terms $\sigma_{w,i}$ and $\sigma_{t_{i,l}}$ have already been defined, $\delta_{l,m}$ is a Kronecker's delta 
and $k_i(\vec{x}_l,\vec{x}_m)$ is the kernel of the GP for instrument $i$, with $\vec{x}_l$ and $\vec{x}_m$ being column vectors of 
columns $l$ and $m$ of the $\mathbf{X}$ matrix. Currently, \codename\ supports the flexible squared-exponential 
kernel in its $D_i$-dimensional form given by
\begin{eqnarray*}
k_i(\vec{x}_l,\vec{x}_m) = \sigma^2_{\mathcal{GP}_i}\exp \left(- \sum^{D_i}_{d=1} \alpha_d (x_{d,l} - x_{d,m})^2\right).
\end{eqnarray*}
Here, $x_{d,l}$ and $x_{d,m}$ are elements ($d,l$) and ($d,m$) of the $\mathbf{X}$ matrix; the $\alpha_d$ 
with $d \in [0,...,D_i]$ and $\sigma_{\mathcal{GP}_i}$ are the hyperparameters of the GP. The $\alpha_d$, on 
one hand, are inverse (squared) length-scale parameters that could be interpreted as the importance of each external 
parameter in our kernel and $\sigma_{\mathcal{GP}_i}$ can be interpreted as the parameter defining the total 
amplitude of the process. This form of the squared-exponential 
kernel has been motivated by the work of \cite{gibson:2012,gibson:2014}, and its success on the modelling 
of systematic trends present in precise transit lightcurves of 
both ground and space-based observatories. {To evaluate the log-likelihood implied 
by this GP, within \codename\ we use the \texttt{george}\footnote{\url{https://github.com/dfm/george}} package 
\citep{george}}.

{Another of the kernels currently supported by 
\codename\ given a vector of inputs $\vec{x}$, is either an exponential GP of the form }
\begin{equation*}
k_i(x_l,x_m) = \sigma^2_{\mathcal{GP},i} \exp \left(-\tau/T_i \right),    
\end{equation*}
where $\tau = |x_l - x_m|$, and the kernel is defined 
by its hyperparameters $\sigma_{\mathcal{GP},i}$ and $T_i$, 
or with an approximate Matern kernel of the form
\begin{eqnarray*}
\nonumber k_i(x_l,x_m) &=& \sigma^2_{\mathcal{GP},i}\left[(1 + 1/\epsilon) e^{-(1-\epsilon)s_i} + (1 - 1/\epsilon)e^{-(1+\epsilon)s_i}\right],
\end{eqnarray*}
with $s_i = \sqrt{3}\tau/\rho_i$, with 
hyperparameters $\sigma_{\mathcal{GP},i}$ and $\rho_i$, and
with $\epsilon$ set to 0.01 \citep[note that as $\epsilon \to 0$, 
this latter kernel converges to a Matern 3/2 kernel; this form and this interpretation was interpreted by][]{celerite}. We also 
consider as a possible kernel the one resulting from 
their multiplication (and which thus only has a common 
amplitude $\sigma^2_{\mathcal{GP},i}$). The latter kernels are introduced and used by \codename\ as 
they are computed using \texttt{celerite}\footnote{\url{https://github.com/dfm/celerite}} \citep{celerite}, 
which provides a very fast means to fit for long term 
trends in datasets with large number of datapoints where 
classical algorithms for GP regression would be too slow. This 
speed improvement, however, comes at a cost: this algorithm can only be applied if the input values are one-dimensional. Also, 
in practice, the $\vec{x}$ vector has to be sorted in ascending 
order (e.g., time is a very good input variable). 

{Finally, \codename\ also supports quasi-periodic 
GP kernels, useful for modelling rotational modulation 
signals in photometric measurements or activity signals 
in radial-velocities.} The 
implied covariance matrix follows the same structure as 
the one defined in equation (\ref{eq:cov}), but for a 
kernel \codename\ either uses a quasi-periodic exp-sine-squared kernel multiplied by a squared 
exponential kernel 
\citep[a model introduced in][for the analysis of photometric and RV data]{haywood:2014}, which 
can be used with any (one) external parameter (e.g., time), the very flexible 
quasi-periodic kernel introduced in \cite{celerite} or the 
more general stochastically-driven damped simple 
harmonic oscillator \citep[SHO, also introduced in][for GP 
regression]{celerite}. The former is of the form
\begin{eqnarray}
\label{eq:qp}
k_i(x_l,x_m) = \sigma^2_{\mathcal{GP}_i}\exp \left(-\alpha_i\tau^2 -\Gamma_i \sin ^2\left[\frac{\pi \tau}{P_{\textnormal{rot},i}}\right]\right),
\end{eqnarray}
and has hyperparameters $\sigma_{\mathcal{GP}_i}$, $\alpha_i$, $\Gamma_i$ and $P_{\textnormal{rot},i}$, with 
the latter defining the period of the quasi-periodic oscillations\footnote{In the notation of \cite{haywood:2014}, $\sigma_{\mathcal{GP}_i}=\eta_1$, $\alpha = 1/2\eta^2_2$, $P_{\textnormal{rot}}=\eta_3$ and $\Gamma = 2/\eta^2_4$. Note here the interpretation of 
$\alpha$ as an inverse squared time-scale.}. {We again use the \texttt{george}\footnote{\url{https://github.com/dfm/george}} package within \codename\ 
to compute the implied log-likelihood this process implies}. On the other hand, the flexible quasi-periodic kernel 
introduced in \cite{celerite} is of the form
\begin{eqnarray}
\label{eq:celerite}
k_i(x_l,x_m) = \frac{B_i}{2+C_i}e^{-\tau/L_i}\left[\cos\left(\frac{2\pi\tau}{P_{\textnormal{rot},i}}\right) + (1+C_i)\right],
\end{eqnarray}
where $B_i$, $C_i$, $L_i$ and $P_{\textnormal{rot},i}$ 
are the hyperparameters of the model, with the latter corresponding to the period of the quasi-periodic oscillations defined by this kernel. The advantage of the latter model is 
that it is faster to perform the computations required for the matrix inversions of the implied covariance matrix as the 
computations within \codename\ for it are generated with \texttt{celerite}. Finally, 
the SHO kernel implemented within \codename\ thanks to its 
\texttt{celerite} implementation \citep{celerite} is of the form 
\[  k_i(x_l,x_m) = S_0\omega_0 Q e^{-\frac{\omega_0 \tau}{2Q}} \left\{
\begin{array}{ll}
      f_1(\omega_0,\tau,Q)\ \ \ \ \ \ \textnormal{for }0 < Q < 1/2,\\[0.5ex]
      f_2(\omega_0,\tau)\ \ \ \ \ \ \ \ \ \ \textnormal{for }Q = 1/2,\\[0.5ex]
      f_3(\omega_0,\tau,Q)\ \ \ \ \ \ \textnormal{for }Q < 1/2,\\[0.5ex]
\end{array} 
\right. \]
where
\begin{eqnarray*}
f_1(\omega_0,\tau,Q) &=& \cosh(\eta \omega_0 \tau) + \frac{1}{2\eta Q}\sinh(\eta \omega_0\tau),\\
f_2(\omega_0,\tau) &=& 2(1 + \omega_0\tau),\\
f_3(\omega_0,\tau,Q) &=& \cos(\eta \omega_0 \tau) + \frac{1}{2\eta Q}\sin(\eta \omega_0\tau),
\end{eqnarray*}
with $\eta = |1 - 1/(4Q^2)|^{1/2}$. This kernel is very 
useful for the modelling of active regions 
in a stellar surface as was noted by \cite{celerite} and also 
used by \cite{Ribas:2018} in the context of modelling stellar activity in radial-velocities.

\begin{table*}
    \centering
    \caption{List of parameters that define each of the 
    noise models for both the radial-velocities (RVs) and photometry for instrument $i$ supported by \codename. Note that in the case in which one global GP is fit to all photometric or RV instruments, only the jitter terms are different for each instrument. For details on each model, see text.}  
    \label{tab:parameters-noise}
    \begin{tabular}{lcl} 
        \hline
        \hline
        Parameter name & Units & Description \\
                \hline
                \hline
        White noise model & & \\
        ~~~$\sigma_{w,i}$ & ppm, m/s or km/s$^1$ & Jitter term. \\
        Multi-dimensional squared-exponential kernel & & \\
        ~~~$\sigma_{w,i}$ & ppm, m/s or km/s$^1$ & Jitter term. \\
        ~~~$\sigma_{\mathcal{GP},i}$ & ppm, m/s or km/s & Amplitude of the GP. \\
        ~~~$\alpha_d$ & --- & Inverse (squared) length-scale parameter for dimension $d$. \\
        Exponential kernel & & \\
        ~~~$\sigma_{w,i}$ & ppm, m/s or km/s$^1$ & Jitter term. \\
        ~~~$\sigma_{\mathcal{GP},i}$ & ppm, m/s or km/s & Amplitude of the GP. \\
        ~~~$T_i$ & --- & Length-scale of the GP process. \\
        Approximate Matern kernel & & \\
        ~~~$\sigma_{w,i}$ & ppm, m/s or km/s$^1$ & Jitter term. \\
        ~~~$\sigma_{\mathcal{GP},i}$ & ppm, m/s or km/s & Amplitude of the GP. \\
        ~~~$\rho_i$ & --- & Length-scale of the GP process. \\
        Exponential times Approximate Matern kernel & & \\
        ~~~$\sigma_{w,i}$ & ppm, m/s or km/s$^1$ & Jitter term. \\
        ~~~$\sigma_{\mathcal{GP},i}$ & ppm, m/s or km/s & Amplitude of the GP. \\
        ~~~$T_i$ & --- & Length-scale of the exponential part of the GP process. \\
        ~~~$\rho_i$ & --- & Length-scale of the (approximate) Matern part of the GP process. \\
        Exp-sine-squared times squared-exponential kernel & & \\
        ~~~$\sigma_{w,i}$ & ppm, m/s or km/s$^1$ & Jitter term. \\
        ~~~$\sigma_{\mathcal{GP},i}$ & ppm, m/s or km/s & Amplitude of the GP. \\
        ~~~$\alpha_i$ & --- & Inverse length-scale of the squared-exponential part of the GP process. \\
        ~~~$\Gamma_i$ & --- & Amplitude of the Exp-sine-squared part of the kernel. \\
        ~~~$P_{\textnormal{rot},i}$ & --- & Characteristic period of the GP. \\
        Celerite quasi-periodic kernel & & \\
        ~~~$\sigma_{w,i}$ & ppm, m/s or km/s$^1$ & Jitter term. \\
        ~~~$B_i$ & ppm, m/s or km/s & Amplitude of the GP. \\
        ~~~$C_I$ & --- & Constant scaling term of the GP. \\
        ~~~$L_i$ & --- & Characteristic time-scale of the GP. \\
        ~~~$P_{\textnormal{rot},i}$ & --- & Characteristic period of the GP. \\
        Stochastic Harmonic Oscillator (SHO) kernel & & \\
        ~~~$\sigma_{w,i}$ & ppm, m/s or km/s$^1$ & Jitter term. \\
        ~~~$S_0$ & --- & Characteristic power of the SHO. \\
        ~~~$\omega_0$ & --- & Characteristic frequency of the SHO. \\
        ~~~$Q$ & --- & Quality factor of the SHO. \\
        \hline
        \hline
    \end{tabular}
    \begin{tablenotes}
      \small
      \item $^1$ If the noise model is defined for the photometry, input is expected to be in parts-per-million (ppm). If defined for the RVs, it is expected to be in the same units as the input RV data (m/s or km/s).
    \end{tablenotes}
\end{table*}

\subsection{Stellar density modelling}
\label{sec:sdmod}
As it is widely known, the scaled semi-major axis and 
the orbital period, parameters obtainable from transiting exoplanet 
lightcurves, allows us to get the stellar density 
of the star hosting the planets which, via Kepler's third law, is given by 
$\hat{\rho}_* = [(3\pi)/(GP_k^2)]\left(a_k/R_*\right)^3$, where 
$G = 6.67408 \times 10^{-11}$ m$^3$ kg$^{-1}$ s$^{-2}$ 
is the gravitational constant, a fact already noted by \cite{sozzetti:2007} to 
constrain the stellar parameters using transit 
lightcurves. However, thanks to the precise measurements 
provided by Gaia Data Release 2 \citep{DR2:2018}, we 
are currently in an era were stellar 
parameters allow us to do the opposite: use the estimated stellar density to precisely constrain the $a_k/R_*$ and $P_k$ of the transiting planets observed in our 
transit lightcurves. Because the periods 
are usually precisely constrained by the periodicity of the 
transits, the stellar density therefore constrains $a_k/R_*$ for transiting planets. Although stating all the benefits that such a constraint provides is outside the scope of this 
work, it is interesting to note that because the transit duration 
along with the ingress and egress times have 
information about the argument of periastron passage 
and the eccentricity of the orbit as well \citep[see, e.g.,][]{winn:2010}, the 
stellar density constraint allows us, in principle, 
to extract part of this information from these observables 
\citep[see, e.g., ][ and references therein]{K:2012,Dawson:2012,K:2014}. 
Similarly, in the prescence of grazing orbits, this 
parametrization allows us to break known degeneracies in the 
transit modelling for a precise estimation of the impact 
parameter --- an effect that also manifestates itself on 
not very well sampled lightcurves \citep{kipping:2010}. Because 
of all these benefits, \codename\ allows to optionally include 
any available estimate on the stellar density as input.

In practice, within \codename\ one can 
incorporate a measured stellar 
density into the modelling in two ways 
if there is only one transiting planet, and in only one way if there are multiple transiting planets. If there is only one 
transiting exoplanet, 
then one way in which we incorporate stellar density 
information is as was already introduced in \cite{Brahm:2018}, where 
the stellar density $\rho_*$ with its associated 
error $\sigma_{\rho_*}$ estimated from the 
modelling of the stellar observables is 
incorporated in the joint modelling as an 
extra dataset/datapoint, $y^{\textnormal{SD}}$. In this case, the probabilistic model for $y^{\textnormal{SD}}$ is 
easy to derive, and is of the form
\begin{eqnarray}
\label{eq:pmodSD}
y^{\textnormal{SD}} \sim \hat{\rho}_{*} + \epsilon_{\textnormal{SD}},
\end{eqnarray}
where $\hat{\rho}_{*} = [(3\pi)/(GP^2)]\left(a/R_*\right)^3$ is 
the model of the stellar density and $\epsilon_{\textnormal{SD}}\sim 
N(0,\sigma_{\rho_*})$, where $\sigma_{\rho_*}$ is the error on our estimate 
for the stellar density. In this case, $y^{\textnormal{SD}}$ is understood as an independant dataset to that of the photometry and radial-velocities and, as such, is easy to incorporate in the joint modelling of the 
data. The 
log-likelihood of the parameters $\vec{\theta} = (a/R_*,P)^T$ implied by the 
probabilistic model in equation \ref{eq:pmodSD} in this case is simply the logarithm of the probability density function of a gaussian distribution 
with mean $(\rho_*-\hat{\rho}_*)$ and variance $\sigma^2_{\rho_*}$. 

The general case of $k$ transiting exoplanets, however, is considerably more complicated to model in the prescence of stellar density information using 
this parametrization, {because each planet would impose a different 
stellar density through $a_k/R_*$, and the star can only have one stellar 
density}. {Because of this, } within \codename\ we have {implemented the possibility to fit directly for the stellar density which, together with the period $P_k$ for each planet, defines through Kepler's law a value of $a_k/R_*$ for each planet}. {This} makes the general case of fitting $k$ transiting exoplanets in the prescence of 
stellar density information a much 
easier problem to solve. Within this 
formulation, the prior is given directly 
on the stellar density parameter, which is 
then used directly as a fitting parameter. \codename\ 
then internally uses the transformation $a_k/R_* = [(\rho_*GP_k^2)/(3\pi)]^{1/3}$ to generate 
the transit models for each individual $k$ 
transiting exoplanet. The advantage of this 
formulation is that this parametrization not 
only constrains the $a_k/R_*$ given observed 
periodic transits, but it also reduces the 
number of fitting parameters by $k-1$ terms. This formulation is also very useful to 
constrain the properties of singly transiting exoplanets.

\subsection{Dynamic, Importance and/or Nested Sampling}
\label{sec:sampling}
{Having defined the probabilistic model \codename\ assumes for the data in 
the previous sections, we now turn to a brief description of the algorithms 
\codename\ uses to both perform posterior sampling in order to obtain the posterior 
distribution of the parameters $\vec{\theta}$ given the data 
$\mathcal{D}$, $p(\vec{\theta} | \mathcal{D})$ \textit{and} to 
estimate bayesian evidences $Z_i = p(\mathcal{D}|M_i)$ for model comparison of each 
model $M_i$ via the posterior odds, $p(M_i|\mathcal{D})/p(M_j|\mathcal{D}) = (Z_i/Z_j)(p(M_i)/p(M_j))$. Currently, \codename\ allows the user to select between 
three possible sampling schemes: nested sampling, importance nested sampling 
and dynamic nested sampling. For detailed overviews, we recommend \cite{MultiNest}, \cite{Multinest2}, and \cite{Buchner:MNEST} for a thorough review of 
nested and importance nested sampling algorithms, and \cite{Buchner:MNEST} and \cite{dynesty} 
for a review on dynamic nested sampling. Our discussion here stems mainly from these 
references.}

{The main idea of classical nested sampling algorithms 
\citep{Skilling:2004,Skilling:2009} is to estimate the 
bayesian evidence of a model, $Z$, by noting that defining the ``prior volume" 
as $X(\lambda) = \int_{\mathcal{L}(\vec{\theta})>\lambda}p(\vec{\theta})d\theta_1 d\theta_2...d\theta_N$, where $\mathcal{L}(\vec{\theta} = p(\mathcal{D}|\vec{\theta})$ is the likelihood 
function, the evidence, which is the integral of the likelihood over the prior 
$p(\vec{\theta})$, can be written as a one-dimensional integral of the 
likelihood over $X$, i.e., 
\begin{eqnarray}
\label{eq:evidence}
Z = \int \mathcal{L}(\vec{\theta})p(\vec{\theta}) d \theta_1d\theta_2...d\theta_N = \int_{0}^{1}\mathcal{L}(X)dX.
\end{eqnarray}
To find the value of $Z$ for the typical case in which the shape of the likelihood function 
is unknown, Monte-Carlo sampling methods are used in order to sequentially shrink the 
prior volume, $X$, by sampling points from the prior $p(\vec{\theta})$. In essence, 
nested sampling algorithms sample $N_\textnormal{live}$ points from this prior and 
sequentially replace in each iteration the live-point with the lowest likelihood by a 
new live point with a larger one. During each iteration, the bayesian evidence is 
updated by a difference $\Delta Z$, and the stopping/convergence criterion is defined 
through a user defined evidence tolerance $\Delta z$ below which the algorithm is said 
to have converged.}

{While classical nested sampling simply reject samples that do not meet the requirement 
of having larger likelihoods than the lowest likelihood sampled by the current set of 
live-points, importance nested sampling uses all the sampled points from the prior, 
assigning different weights to each sampled value, which in turn leads to a better usage of the sampling and consequently a much faster 
convergence \cite[see][for details]{INS,Multinest2}. Within \codename\ we have incorporated both of these sampling schemes through 
the \texttt{MultiNest} algorithm \citep{MultiNest,Multinest2} via the \texttt{PyMultiNest}\footnote{\url{https://github.com/JohannesBuchner/PyMultiNest}} 
package \citep{PyMultiNest}. This algorithm is very efficient especially for multi-modal 
distributions as it encloses the live-points within ellipsoids, which can eventually be 
disconnected when different modes are found, allowing to accomodate the algorithm for very 
complicated shapes of the posterior distribution \cite[see, e.g., examples in][]{Multinest2}. 
Within \codename\ we use the default tuning parameters of \texttt{MultiNest}, which as 
shown in \cite{Multinest2} allow to tackle a wide range of posteriors including 
distributions much more complicated than the ones one might expect when 
fitting photometry and radial-velocityies of exoplanets. In particular, the default value 
within \codename\ for $\Delta z$ is the default set in \texttt{MultiNest} which is $0.5$ 
(it is important to notice that this is \textit{not} the error on the evidence --- this 
is typically more than an order of magnitude smaller than this stopping value). The 
number of live-points is set to a default of 1000, but it can be defined by the user.}

{One of the main drawbacks of nested sampling algorithms such as \texttt{MultiNest} 
is that because they are focused on evidence calculations, the posterior distribution of 
the parameters $p(\vec{\theta}|\mathcal{D})$ is only a by-product of the procedure, 
which in turn might not explore as efficiently as possible the parameter space. The main 
reason for the inability of nested samplers to focus on posterior samples is because the 
number of live points $N_\textnormal{live}$ is kept constant, and thus the prior volume 
shrinkage rate (i.e., the shrinking rate of $X$) is always the same. To solve 
this, \cite{Higson:2017} proposed to dynamically change the number of live-points in 
order to change the focus of the algorithm during the run. This sampling scheme, 
called ``dynamic`` nested sampling, has been implemented by \cite{dynesty} through 
the \texttt{dynesty}\footnote{\url{https://github.com/joshspeagle/dynesty}} library, which we also incorporate within \codename. Here, however, 
the convergence criteria is much more involved to define as it is not solely evidence based, 
but also based on how much of the posterior distribution has been explored. Within 
\codename\ we use the default criterion described in Section 3.4 of \cite{dynesty}, 
which is the one implemented in \texttt{dynesty}.}

{We refer the reader to \cite{Buchner:MNEST} for a discussion 
on the advantages of dynamic nested sampling over ``regular" nested sampling, but at 
first order, we recommend selecting to use \texttt{MultiNest} for (importance) nested sampling or \texttt{dynesty} for dynamic nested sampling within \codename\ having the number of 
dimensions (i.e., the number of parameters 
to be fitted) in mind: in our experience, MultiNest works well up 
to $\sim 20-$dimensional problems. For this order of magnitude 
problems and/or larger, users should use \texttt{dynesty} which 
provides algorithms specifically designed to tackle such large-dimensional problems. Both algorithms 
offer multi-threading options that one can use 
within \codename\ (via OpenMPI for MultiNest and via 
internal \texttt{python} multi-threading options for 
\texttt{dynesty}), allowing thus to scale the speed 
of convergence on multiple-core machines for large 
high-dimensional problems.}

\subsection{Joint modelling, model selection and parameter estimation}
{Using the sampling methods described in Section \ref{sec:sampling} 
and the probabilistic models described in previous sections, \codename\ allows to perform 
a} joint modelling of transits and radial-velocities{. W}e allow for all the 
possibilities: both transiting and non-transiting systems, and systems 
that can and cannot be detected via radial-velocities 
(which, of course, include the possible fitting of 
only transits and only radial-velocities). {In practice, this implies we 
need to} define the priors for our parameters (which are user-defined) and the likelihood function, which in our case is easy to compute as the sum of the individual 
log-likelihoods defined for the photometry and radial-velocities, which have the form 
defined in equation (\ref{eq:like}) plus the likelihood for the stellar density defined 
in the previous sub-section {if one uses this as a datapoint. This latter term is of course not used if the stellar density is used as a parameter in the fit.}

It is important to discuss the advantages and caveats that nested samplers offer 
for both model selection and parameter estimation, which is one of 
the main differences between \texttt{juliet} and other {codes} for 
analyzing photometric and radial-velocity measurements for the 
interpretation of exoplanetary signals. As it was briefly discussed 
above, nested sampling algorithms are specifically made for the estimation of 
bayesian evidences of different models given the data, $Z_i$, which 
in turn provide us with a {useful} tool for model selection. However, {it 
is important to highlight that model evidences are heavily impacted by the priors used 
to estimate them and, thus, special care must be taken when selecting them. Also of 
importance are the errors on the evidences quoted by nested sampling algorithms. 
According to the recent work of \citep{Nelson:2018},} at least 
for cases in which one is performing both parameter estimation and 
model selection on radial-velocity only datasets {they} seem to be 
underestimated, and thus care must be taken on this front 
when comparing evidences between different models. 
In our experience with \texttt{juliet}, however, having transits and 
radial-velocities being fitted simultaneously significantly 
leverages this problem, as the periodic nature of transits 
precisely constrain the ephemerides of the exoplanetary signals. 
Although providing an 
extensive analysis such as the one of \cite{Nelson:2018} when it comes 
to model comparison and selection in datasets with both transits and 
radial-velocities is out the scope of this work, this possibility 
should still be kept in mind. We thus echo the recommendations in 
\cite{Nelson:2018}: to run a handful of \texttt{juliet} runs when 
comparing model evidences in different cases to check for consistency 
between the estimates of the bayesian evidences provided by the 
nested sampling algorithm under use.

In terms of parameter estimation, nested samplers offer a very 
important advantage: given a thorough exploration of the 
prior volume is needed in order to compute model evidences, 
nested sampling algorithms are very efficient at thoroughly 
exploring the parameter space, and thus, on the search for 
not only the global optimum but also for local minima which 
might be true solutions of a possibly multi-modal posterior 
distribution. Given proper priors, this provides 
a very efficient venue to search for the optimal parameters 
during the posterior distribution exploration simultaneously. This 
property is unlike Markov-Chain Monte Carlo (MCMC) algorithms such as the classical Metropolis-Hastings algorithms 
\citep[see, e.g., the discussion in][]{Tak:2018} or of widely 
used fast algorithms such as \texttt{emcee} \citep{emcee} which 
require initial guesses of the parameters to be made in order 
to attain convergence in a reasonable amount of time. This 
is especially important for posterior exploration of 
models using GPs, for which a search of optimal parameters 
before posterior exploration is usually a though task to 
optimize in itself. Of course, the larger the prior volume 
the slower the convergence of nested sampling algorithms 
as well --- however, the multi-threading capabilities of the 
nested sampling algorithms used within \texttt{juliet}, 
{can help in leveraging this issue} for typical 
problems in exoplanetary science. 

\section{Modelling planetary systems with \codename}
\label{sec:tests} 

{Having defined the probabilistic models and sampling 
capabilities \codename\ offers in the past sections, here we} 
showcase some of the features \codename\ offers for the analysis 
of transiting and non-transiting exoplanetary systems. {In 
order to do this}, we perform a detailed analysis of two 
interesting systems. First, we perform a thorough 
analysis of the K2-140b system, which is a hot-Jupiter that 
has been claimed to have a significant eccentricity in 
\cite{Giles:2018}. This claim, however, has been put into question 
by the analysis in \cite{Korth:2019}. We believe 
this system is interesting for us to showcase the power \codename\ 
has for providing a quantitave view not only into the question of 
whether a given dataset supports or not the presence of a property 
such as a significant eccentricity, but also to illustrate the 
interpretation of dilution factors, how \codename\ handles whether 
GPs are needed to account for additional systematics in the data and 
how stellar density priors impact on the analysis of a system. After 
this analysis, we perform an analysis of the K2-32 multi-planet system, 
whose radial-velocities have been analyzed in \cite{K2-32} but for which 
a joint analysis has not been presented in the literature so far. For 
the analysis of the K2-140 system we use \texttt{MultiNest}, as the number 
of parameters in all of our analyses are on the order of 20 free parameters. 
For the analysis of the K2-32 system we use \texttt{dynesty}, as in 
this case the number of free parameters is larger (30). In what follows and 
following \cite{Trotta:2008}, we consider $\Delta \ln Z = 2$ (or posterior 
odds of $\approx 7:1$, assuming equiprobable models) as a threshold between 
weak and moderate evidence that one model is preferred over the other, 
whereas $\Delta \ln Z = 5$ (or posterior odds of $\approx 150:1$ assuming 
equiprobable models) as strong evidence of one model over the 
other. {We have run all the examples below several times 
and have found that the (log-)evidence estimates have errors $\lesssim$ 0.1; we thus don't quote values for 
the evidences smaller than this.}

\subsection{Analyzing the K2-140 system with \codename}
\label{sec:k2140}
The data used here for K2-140 considers (1) the K2 photometry, (2) follow-up photometry from the Las Cumbres Observatory Global Telescope (LCOGT) 
Network, (3) radial-velocities from CORALIE and HARPS \citep[photometric 
and radial-velocity data already introduced in][]{Giles:2018} and (4) radial-velocities from the FIES 
instrument, introduced in \cite{Korth:2019}. This dataset, along with 
the scripts used to generate the analysis to be presented in this 
Section {(including sampling parameters such as, e.g., the number of live-points)} are made available in the \texttt{juliet} Github wiki page\footnote{\url{https://github.com/nespinoza/juliet/wiki}}.

\subsubsection{Photometric analyses with \codename}
We first analyze the photometry of K2-140b using \codename. First of 
all, we must consider the fact that the K2 photometry for K2-140 was 
obtained in the long-cadence mode of K2 photometry, which if not 
accounted for can give rise to biases in the retrieved transit 
parameters \citep{kipping:2010}. As such, we account for this 
effect within \codename\ by supersampling the lightcurve using the 
\textit{Kepler} exposure time (0.020434 days), and super-sampling 
$N=20$ datapoints at each K2 time-step, averaging these datapoints in 
order to generate our model lightcurve. Inside \codename, this process 
is easily defined by flags the user can give before starting  a 
\codename\ run. In practice, the supersampling is actually performed by 
\texttt{batman} \citep{batman:2015}, which 
is the transit modelling tool \codename\ uses to generate the transit 
models. In addition, following \cite{EJ:2016} we use a quadratic 
limb-darkening law for the precise K2 photometry and a linear law 
for the less precise LCOGT photometry; these are also easily defined 
via flags within \codename.

\subsubsection{Dilution factors}
\label{sec:dilution}
As a first fit to the data, we try a white-noise analysis to the K2 
photometry and LCOGT photometry simultaneously assuming a circular 
orbit and letting all the parameters have the wide uninformative 
priors presented in Table \ref{tab:priors-photometry}. 
In order to decide whether we should include the dilution factors and 
on which instruments, we try turning off the dilutions of each 
instrument. The retrieved log-evidence indicates that the best model 
of those is one in which we fix the dilution to 1 for both 
instruments ($\Delta \ln Z > 2$ for this model when compared against 
all the other models). All the other models are statistically 
indistinguishable from one another (all have $\Delta \ln Z < 2$ 
between themselves). One might, however, have 
good reason to believe a-priori that the best would be to leave the 
dilution factor for the K2 photomery as a free parameter in the fit, 
as the \textit{Kepler} pixels are relatively large ($\sim$4'') and, 
indeed, the aperture used to extract the K2 photometry might include light 
from nearby sources that could be diluting the transit. 

\begin{figure}
   \includegraphics[height=0.75\columnwidth]{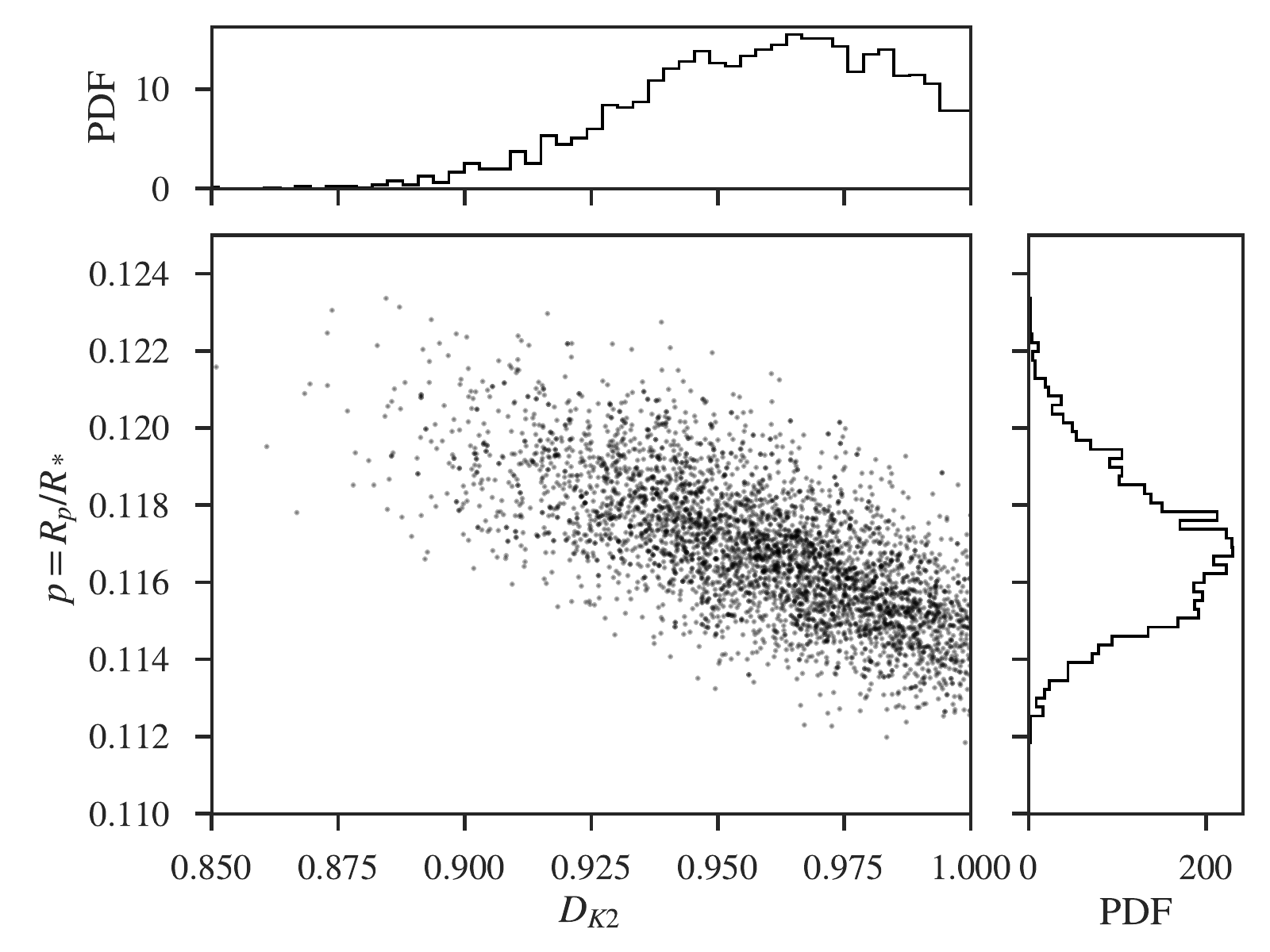}
    \caption{Posterior distribution of the dilution factor for the K2 
    photometry versus the 
    planet-to-star radius ratio for K2-140b. Note how the correlation 
    between the parameters enlarges the posterior distribution of the 
    planet-to-star radius ratio.}
    \label{fig:dfactor}
\end{figure}

In Figure \ref{fig:dfactor} we present the posterior distribution of 
the dilution factor for the K2 photometry versus the planet-to-star 
radius ratio for the case in which the LCOGT photometry dilution factor 
is fixed to 1 (which is a reasonable assumption as the aperture used 
to obtain this photometry is on the order of arcseconds), and in which 
it can be seen how the correlation between these 
two parameters enlarges the posterior distribution of the latter making 
its marginal posterior more uncertain. It is interesting to interpret 
what this dilution factor \textit{means} in light of our discussion in 
Section \ref{sec:photmod}. As it was discussed in that section, we 
can interpret the dilution factor as a measure of the light that is 
being added to the aperture by other sources; i.e., 
\begin{eqnarray*}
\sum_n F_n/F_T = \frac{1-D}{D}.
\end{eqnarray*}
In our case, $\sum_n F_n/F_T = 4.2^{+3.1}_{-2.6} \%$. Multiplying this 
by -2.51 and taking the logarithm base 10, we find that if this dilution 
was indeed made by a nearby star, this would imply a star with a 
delta-magnitude $\Delta m = 3.5^{+1.0}_{-0.6}$ fainter than the 
target star; in general, $n$ sources of magnitudes $\Delta m = 2.51\log_{10}(n) + 3.5^{+1.0}_{-0.6}$ can produce such a dilution. 
Looking at nearby stars around K2-140 (i.e., stars within $\sim 4$ 
\textit{Kepler} pixels or $\sim 16$ arcseconds to it) in Gaia DR2 
\citep{DR2:2018}, we see that there is only one source (Gaia ID: 3579426058019822080) that 
\textit{might} add flux to the K2 aperture used to obtain the 
photometry of K2-140. However, this source is much too 
faint ($\Delta G = 8.7$ fainter than the target star, which has 
$G = 12.48$) to produce this level of dilution. Given that the 
field does not seem to be especially crowded according to the 
Gaia detections around 10 arcminutes of the target (441 sources 
down to $G = 21$, or around 1.4 sources per arcminute$^2$), it seems 
very unlikely that there are undetected sources blended with K2-140 
that could give rise to the level of dilution implied by our K2 
observations when leaving this parameter be free in the fit. {A close exploration 
of the posterior distribution of the parameters reveals that the increase in the width of 
the marginal posterior distribution of the planet-to-star radius ratio is due to the 
limb-darkening coefficients, which in this case have a strong correlation with both this 
parameter and the dilution factor (which makes sense, as those parameters heavily influence 
on the shape of the transit lightcurve). Based on our evidence calculation and in our posterior 
information regarding close-by sources, we} thus 
decide to use the simpler model suggested by our log-evidence calculations, 
in which the dilution factors of both of our instruments are set to 
unity in what follows.

\begin{table*}
    \centering
    \caption{Priors used in the analysis of the photometry for K2-140b. Here $p=R_p/R_*$ 
    and $b=(a/R_*)\cos(i_p)$, where $R_p$ is the planetary 
    radius, $R_*$ the stellar radius, $a$ the semi-major axis of the 
    orbit and $i_p$ the inclination of the planetary orbit with respect 
    to the plane of the sky. $e$ and $\omega$ are the eccentricity and 
    argument of periastron of the orbits. $\mathcal{N}(\mu,\sigma^2)$ represents a normal distribution of mean $\mu$ and variance $\sigma^2$. $\mathcal{U}(a,b)$ represents a uniform distribution between $a$ and $b$. $\mathcal{J}(a,b)$ represents a Jeffrey's 
    prior (i.e., a log-uniform distribution) between $a$ and $b$.}  
    \label{tab:priors-photometry}
    \begin{tabular}{lccl} 
        \hline
        \hline
        Parameter name & Prior & Units & Description \\
                \hline
                \hline
        Parameters for \planetnameb & & \\
        ~~~$P_b$ &$\mathcal{N}(6.5693,0.01^2)$ & days & Period of \planetnameb. \\
        ~~~$t_{0,b}$ &$\mathcal{N}(2457588.28380,0.01^2)$ & days & Time of transit-center for \planetnameb. \\
        ~~~$a_{b}/R_*$ &$\mathcal{U}(1,30)$ & stellar radii & Scaled semi-major axis for \planetnameb. \\
        ~~~$r_{1,b}$ &$\mathcal{U}(0,1)$ & --- & Parametrization$^{1}$ of \cite{Espinoza:2018} for $p$ and $b$ for \planetnameb. \\
        ~~~$r_{2,b}$ &$\mathcal{U}(0,1)$ & --- & Parametrization$^{1}$ of \cite{Espinoza:2018} for $p$ and $b$ for \planetnameb. \\
        \hline
        Parameters for K2 photometry & & \\
        ~~~$D_{\textnormal{K2}}$ & $\mathcal{U}(0,1)$ & --- & Dilution factor for K2. \\
        ~~~$M_{\textnormal{K2}}$ &$\mathcal{N}(0,0.1^2)$ & relative flux & Relative flux offset for K2. \\
        ~~~$\sigma_{w,\textnormal{K2}}$ &$\mathcal{J}(0.1,500^2)$ & relative flux (ppm) & Extra jitter term for K2 lightcurve. \\
        ~~~$q_{1,\textnormal{K2}}$ &$\mathcal{U}(0,1)$ & --- & Quadratic limb-darkening parametrization$^{3}$ \citep{kipping:2013}. \\
        ~~~$q_{2,\textnormal{K2}}$ &$\mathcal{U}(0,1)$ & --- & Quadratic limb-darkening parametrization$^{3}$ \citep{kipping:2013}. \\  
        \hline
        Parameters for LCOGT photometry & & \\
        ~~~$D_{\textnormal{LCOGT}}$ & $\mathcal{U}(0,1)$ & --- & Dilution factor for LCOGT. \\
        ~~~$M_{\textnormal{LCOGT}}$ &$\mathcal{N}(0,0.1^2)$ & relative flux & Relative flux offset for LCOGT. \\
        ~~~$\sigma_{w,\textnormal{LCOGT}}$ &$\mathcal{J}(0.1,5000^2)$ & relative flux (ppm) & Extra jitter term for LCOGT lightcurve. \\
        ~~~$q_{1,\textnormal{LCOGT}}$ &$\mathcal{U}(0,1)$ & --- & Linear limb-darkening coefficient for the LCOGT photometry. \\
        \hline
        \hline
    \end{tabular}
    \begin{tablenotes}
      \small
      \item $^1$ To perform the transformation between the $(r_1,r_2)$ plane and the $(b,p)$ plane, we performed the 
      transformations outlined in \cite{Espinoza:2018} depending on the 
      values of $r_{1}$ and $r_{2}$: with $p_l=0$ and $p_u=1$, if $r_{1}>A_r = (p_u-p_l)/(2 + p_l + p_u)$, then $(b,p) = ([1+p_l][1+(r_{1}-1)/(1-A_r)], (1-r_{2})p_l + r_{2}p_u)$. If 
      $r_{1}\leq A_r$, then $(b,p) = ([1+p_l] + \sqrt{r_{1}/A_r}r_{2}(p_u-p_l), p_u + (p_l-p_u)\sqrt{r_{1}/A_r}[1-r_{2}])$.
      \item $^3$ To transform from the $(q_1,q_2)$ plane to the plane of 
      the quadratic limb-darkening coefficients, $(u_1,u_2)$, we use 
      the transformations outlined in \cite{kipping:2013} for this 
      law $u_1 = 2\sqrt{q_1}q_2$ and $u_2=\sqrt{q_1}(1-2q_2)$.
    \end{tablenotes}
\end{table*}

\subsubsection{Gaussian Processes}
\begin{figure*}
   \includegraphics[height=0.8\columnwidth]{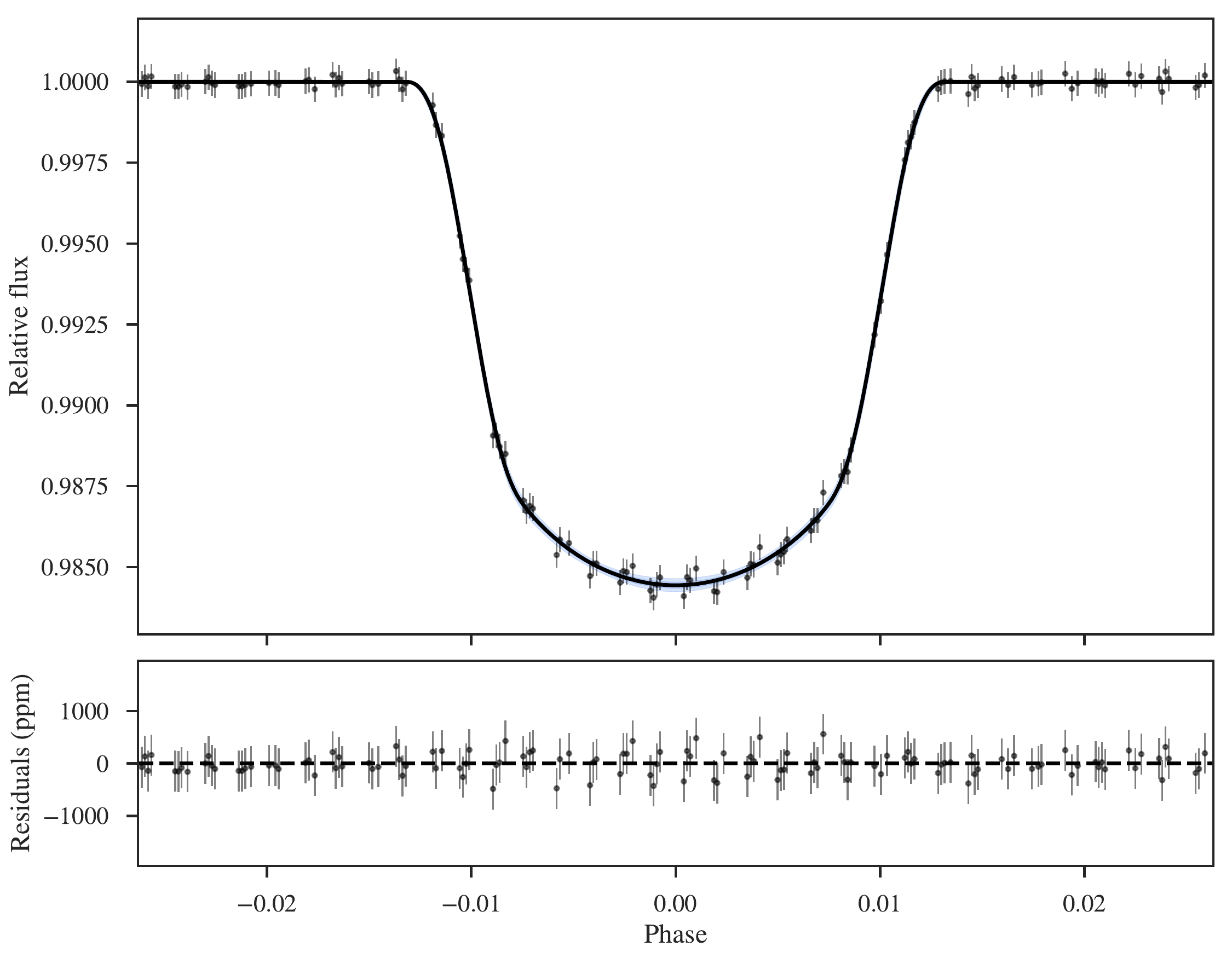}
   \includegraphics[height=0.8\columnwidth]{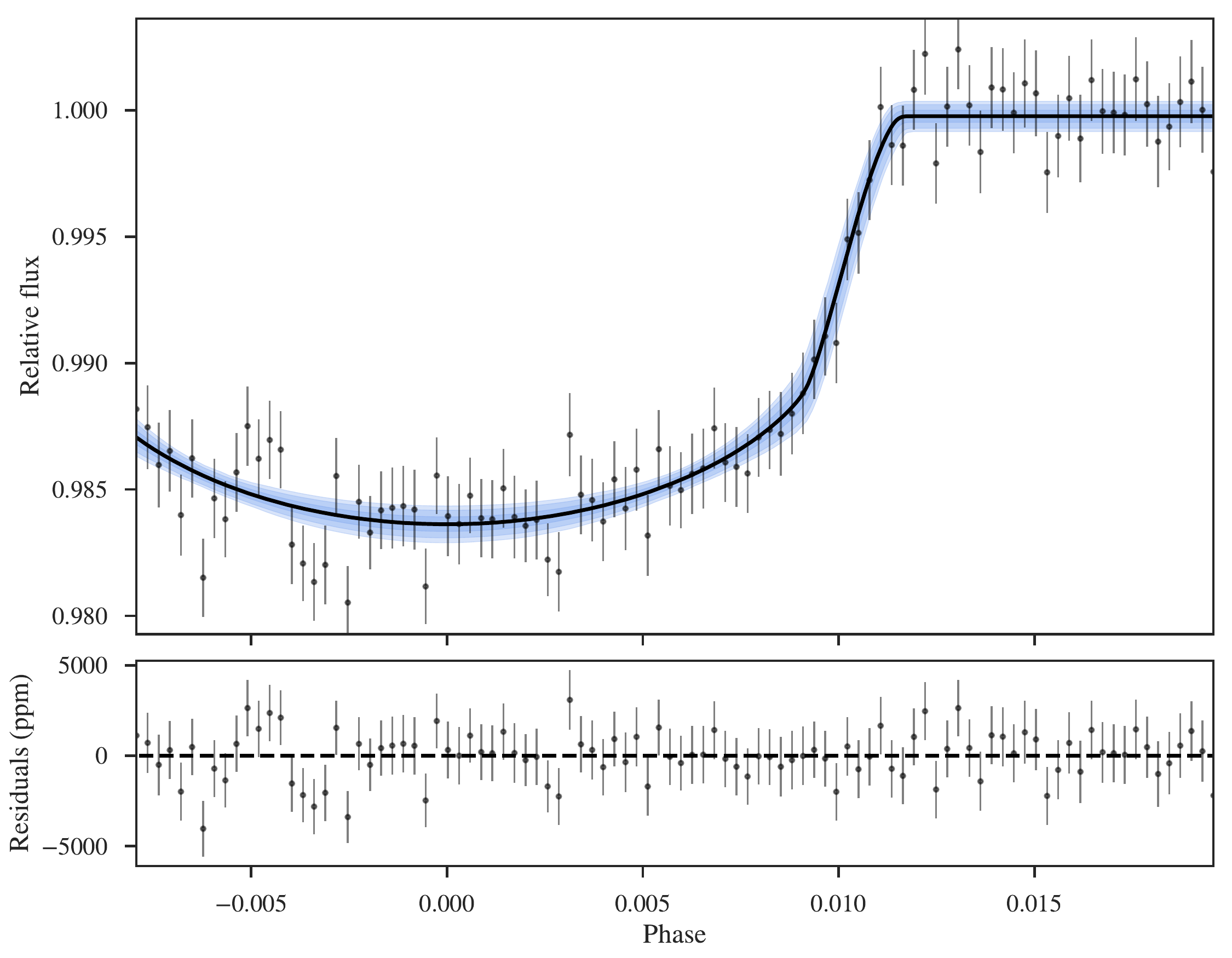}
    \caption{Lightcurve fits of our \codename\ run on the photometry for K2-140b 
    (left, K2 photometry; right, LCOGT photometry). The errorbars include the 
    fitted photometric jitter terms added in quadrature. Solid black lines show 
    our best-fit models; blue bands denote 68, 95 and 99\% posterior 
    credibility bands for our best-fit models.}
    \label{fig:lcfits}
\end{figure*}

As discussed in Section \ref{sec:photmod}, \codename\ is able to fit 
GPs given any external parameters to each photometric instrument in 
order to account for possible systematic effects. The K2 photometry 
presented in \cite{Giles:2018} has already been detrended and as 
such we do not bother to try a GP on that dataset, but we might 
believe that the LCOGT photometry might need some systematics 
correction. To test this, we model any systematic trends in the LCOGT 
photometry with a simple squared-exponential kernel in time, setting 
a prior of $\sigma_{\mathcal{GP},\textnormal{LCOGT}} \sim \mathcal{J}(1,10^4)$ and 
$\alpha_{\hat{t},\textnormal{LCOGT}} \sim \mathcal{J}(0.01,10)$ for this instrument, where $\hat{t}$ is the 
time substracted by its mean and divided by its standard deviation 
(i.e., we standarize the time variable). These parameters are added 
to the priors defined in Table \ref{tab:priors-photometry}, and as 
was discussed in Section \ref{sec:dilution} we fix the dilution 
factors to 1 for both instruments.

The resulting evidences for the GP and no-GP fits with \codename\ 
reveal that the evidence for both models are statistically indistinguishable 
($\Delta \ln Z < 1$); however, the no-GP model's evidence is actually larger 
than the GP model. As such, we decide to use the model without this GP 
in time. We show the resulting fits of this photometric \codename\ 
analysis in Figure \ref{fig:lcfits}. 

\subsubsection{Radial-velocity analyses with \codename}
\label{sec:rvtest}
We now analyze the radial-velocities with \codename, in order to showcase the 
features our code provides for the analysis of this kind of data. We 
perform two analyses: one in which we first consider ourselves agnostic as 
to whether there is a planetary signal in the radial-velocity data or not, 
and another one in which we use the ephemerides obtained for our photometric 
analysis in the previous sub-section in order to search for evidence of 
eccentricity in the radial velocities. 

\begin{table*}
    \centering
    \caption{Priors used in the ``agnostic" radial-velocity analysis of the \starname\ system using \codename\, on the search for evidence 
    in the RVs for radial-velocity variations on \starname\ due to \planetnameb. A circular orbit is assumed.}  
    \label{tab:priors-agnostic}
    \begin{tabular}{lccl} 
        \hline
        \hline
        Parameter name & Prior & Units & Description \\
                \hline
                \hline
        Parameters for \planetnameb & & \\
        ~~~$P_b$ &$\mathcal{J}(0.1,100)$ & days & Period of \planetnameb. \\
        ~~~$t_{0,b}$ &$\mathcal{U}(2457803,2457905)$ & days & Time of transit-center for \planetnameb. \\
        ~~~$K_{b}$ &$\mathcal{U}(0,1000)$ & m/s & Radial-velocity semi-amplitude for \planetnameb. \\
        \hline
        RV parameters & & \\
        ~~~$\mu_{\textnormal{CORALIE}}$ &$\mathcal{N}(1220,50^2)$ & m/s & Systemic velocity for CORALIE. \\
        ~~~$\sigma_{w,\textnormal{CORALIE}}$ &$\mathcal{J}(0.1,100)$ & m/s & Extra jitter term for CORALIE. \\
        ~~~$\mu_{\textnormal{HARPS}}$ &$\mathcal{N}(1240,50^2)$ & m/s & Systemic velocity for HARPS. \\
        ~~~$\sigma_{w,\textnormal{HARPS}}$ &$\mathcal{J}(0.1,100)$ & m/s & Extra jitter term for HARPS. \\
        ~~~$\mu_{\textnormal{FIES}}$ &$\mathcal{N}(1215,50^2)$ & m/s & Systemic velocity for FIES. \\
        ~~~$\sigma_{w,\textnormal{FIES}}$ &$\mathcal{J}(0.1,100)$ & m/s & Extra jitter term for FIES. \\      
        \hline
        \hline
    \end{tabular}
\end{table*}

\begin{figure*}
   \includegraphics[height=0.55\columnwidth]{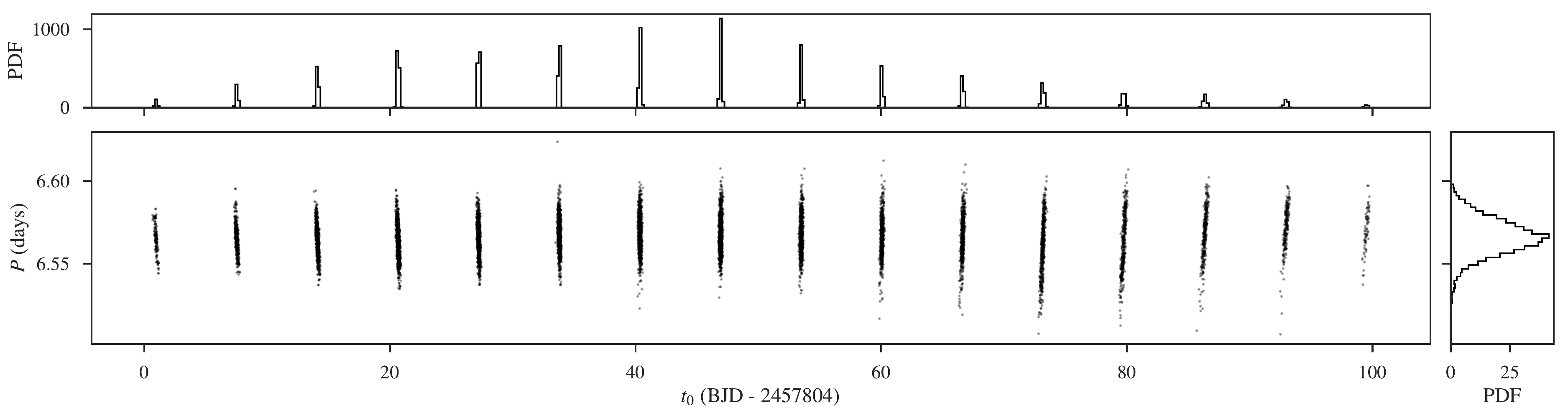}
    \caption{Posterior distribution of the time of transit center and period 
    of the orbit of our ``agnostic" analysis. This distribution showcases one 
    of the features of nested samplers: their ability to sample from a posterior 
    with multiple-modes.}
    \label{fig:rvagnostic}
\end{figure*}

For our ``agnostic" analysis, we perform a \codename\ run with 
the priors presented in Table \ref{tab:priors-agnostic}. As a null model, we 
perform an analysis using the same priors for all the parameters but the jitters, 
whose priors we set from 0.1 to 1000 m/s, and the planetary elements, where 
we fix the semi-amplitude to zero. This analysis provides ample evidence 
for a planetary signal when 
compared against the null model: the evidence for the planetary model is 
$\Delta \ln Z = 10$ larger than for the null model (i.e., the planetary model 
is 4 orders of magnitude larger than the null model). In turn, the period and 
time of transit center(s) we obtain with our analysis, 
$t_0=2457588.40\pm 0.58$ (obtained substracting 48 times the best-fit 
period to match the time-of-transit center of the K2 observations), 
$P=6.566 \pm 0.015$, whose joint posterior distribution is presented in 
Figure \ref{fig:rvagnostic}, perfectly agrees with the ephemerides observed 
in the transit data, for which we obtained 
$t_0 = 2457588.28379 \pm 0.00026$ and $P = 6.569301 \pm 0.000029$. 
It is important to notice that this latter distribution is multi-modal 
for obvious reasons: a time of transit plus or minus $nP$ is also a solution 
to the problem. {However, not all the modes are equally precise: 
the most precise in terms of the time-of-transit center is, in fact, the one 
in the middle of the observing window. Modes farther away from this get 
more and more horizontal in the $(t_0, P)$-plane, meaning the correlations between $t_0$ and 
$P$ is stronger. This makes sense as to predict the next time-of-transit center 
one has to add $n$ times the period $P$, making future and past predictions of the 
time-of-transit center much more imprecise and correlated with $P$.} This in 
turn showcases one of the features that \codename\ provides thanks to the 
nested sampling algorithms it uses: it easily handles multi-modal distributions. 

Having ample evidence for a planetary signal in the radial-velocities, 
we now turn to our analysis on the search for evidence of eccentricity in the 
radial-velocity dataset only. For this, we use the same priors as the ones 
presented in Table \ref{tab:priors-agnostic}, but now use as priors on 
the ephemerides of the orbit the period and time-of-transit center 
found with our photometric-only analysis. In addition to this, we fit for 
$\mathcal{S}_{1} \sim \mathcal{U}(-1,1)$ and $\mathcal{S}_{2} \sim \mathcal{U}(-1,1)$ in the eccentric case, and set the eccentricity to 
0 in the circular case. 

\begin{figure*}
   \includegraphics[height=2.0\columnwidth]{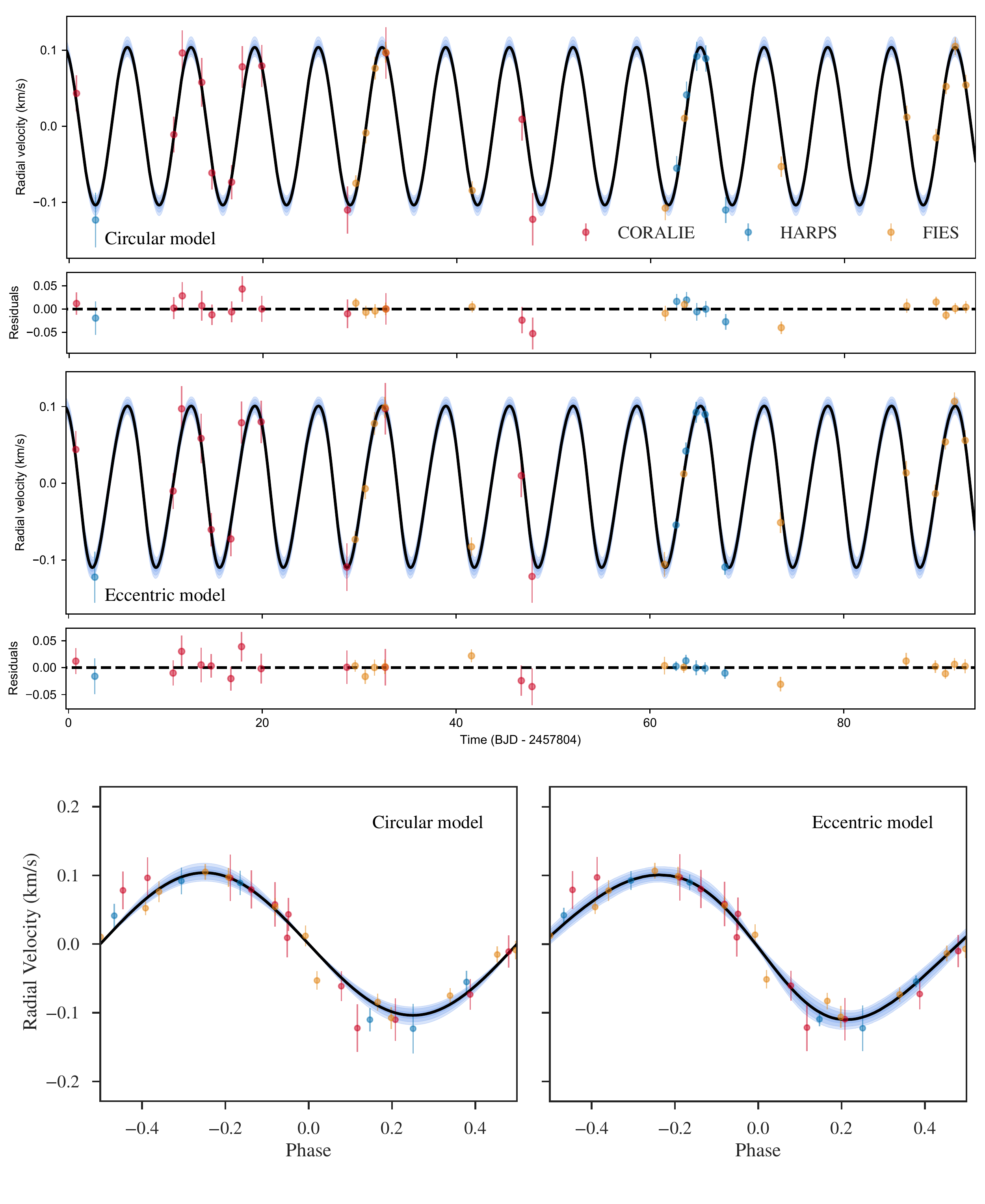}
    \caption{Radial-velocity fits for K2-140b both using a circular 
    (upper panels and lower left panel) and an eccentric (upper panels 
    and lower right panel) model. According to the analysis of the 
    bayesian evidences, both models are indistinguishable from 
    one another (see text).}
    \label{fig:rvfits}
\end{figure*}

The results of our analysis are presented in Figure \ref{fig:rvfits}. 
In terms of the retrieved evidences, our results are ambiguous: the 
eccentric model has a log-evidence only $\Delta \ln Z = 0.20$ larger 
than the circular model, which makes both models statistically indistinguishable. From Figure \ref{fig:rvfits} we can visually see 
that this reflects exactly what we observe when comparing the circular 
and eccentric fits given the data at hand --- there is no evident 
improvement on the fits when inspecting either the residuals of the 
circular or eccentric model fit. Using only the radial-velocities, thus, 
it is impossible to determine if the system is indeed eccentric or not: 
given this data only and the priors used, both the circular and 
eccentric models are indistinguishable.

\subsubsection{Joint-analyses with \codename}
We now turn to the final analysis of the K2-140 dataset, on which we 
perform joint analyses of the photometry and radial-velocities with 
\codename\ in order to showcase the impact this kind of analysis have 
on helping constrain the planetary parameters from this data. We perform 
two joint analyses here: one in which we ignore the stellar parameters, 
and one in which we incorporate the stellar density information into our modelling.

For our first set of analyses, we perform a joint analysis of the 
photometry and radial-velocities with \codename\ using the priors 
defined in Table \ref{tab:joint}. For the priors in the ephemerides 
of our joint fit, we inflated the uncertainties found from our 
photometry-only analysis. Our joint fitting of the data reveals 
plots very similar to the ones already presented in Figures 
\ref{fig:lcfits} and \ref{fig:rvfits}; however, the difference between 
the evidences of both the circular and eccentric fits are slightly 
different: we obtain a $\ln \Delta Z = 0.6$ in favor of the eccentric 
model. Although formally both models are indistinguishable, 
it is interesting to see the slight increase in the evidence in this case 
when compared with the radial-velocity only dataset, which might be a 
product of the increased information the transit photometry gives to 
the final orbits. This, however, has a null impact on the actual posterior 
distributions of the eccentricity and argument of periastron, which is 
virtually the same as the ones we observe for the case in which we fit 
for the radial-velocity dataset only.

\begin{table*}
    \centering
    \caption{Priors used in our joint analysis of the \starname\ system using \codename. }  
    \label{tab:joint}
    \begin{tabular}{lccl} 
        \hline
        \hline
        Parameter name & Prior & Units & Description \\
                \hline
                \hline
        Parameters for \planetnameb & & \\
        ~~~$P_b$ &$\mathcal{N}(6.5693,0.0001^2)$ & days & Period of \planetnameb. \\
        ~~~$t_{0,b}$ &$\mathcal{N}(2457588.284,0.001^2)$ & days & Time of transit-center for \planetnameb. \\
        ~~~$a_{b}/R_*$ &$\mathcal{U}(1,30)$ & stellar radii & Scaled semi-major axis for \planetnameb. \\
        ~~~$r_{1,b}$ &$\mathcal{U}(0,1)$ & --- & Parametrization$^{1}$ of \cite{Espinoza:2018} for $p$ and $b$ for \planetnameb. \\
        ~~~$r_{2,b}$ &$\mathcal{U}(0,1)$ & --- & Parametrization$^{1}$ of \cite{Espinoza:2018} for $p$ and $b$ for \planetnameb. \\
        ~~~$\mathcal{S}_{1,b} = \sqrt{e_b}\sin \omega_b$ &$\mathcal{U}(-1,1)$ & --- & Parametrization$^{2}$ for $e$ and $\omega$ for \planetnameb. \\
        ~~~$\mathcal{S}_{2,b} = \sqrt{e_b}\cos \omega_b$ &$\mathcal{U}(-1,1)$ & --- & Parametrization$^{2}$ for $e$ and $\omega$ for \planetnameb. \\
        ~~~$K_{b}$ &$\mathcal{U}(0,1000)$ & m/s & Radial-velocity semi-amplitude for \planetnameb. \\
        \hline
        Parameters for K2 photometry & & \\
        ~~~$D_{\textnormal{K2}}$ & 1 (fixed) & --- & Dilution factor for K2. \\
        ~~~$M_{\textnormal{K2}}$ &$\mathcal{N}(0,0.1^2)$ & relative flux & Relative flux offset for K2. \\
        ~~~$\sigma_{w,\textnormal{K2}}$ &$\mathcal{J}(0.1,500^2)$ & relative flux (ppm) & Extra jitter term for K2 lightcurve. \\
        ~~~$q_{1,\textnormal{K2}}$ &$\mathcal{U}(0,1)$ & --- & Quadratic limb-darkening parametrization$^{3}$ \citep{kipping:2013}. \\
        ~~~$q_{2,\textnormal{K2}}$ &$\mathcal{U}(0,1)$ & --- & Quadratic limb-darkening parametrization$^{3}$ \citep{kipping:2013}. \\  
        \hline
        Parameters for LCOGT photometry & & \\
        ~~~$D_{\textnormal{LCOGT}}$ & 1 (fixed) & --- & Dilution factor for LCOGT. \\
        ~~~$M_{\textnormal{LCOGT}}$ &$\mathcal{N}(0,0.1^2)$ & relative flux & Relative flux offset for LCOGT. \\
        ~~~$\sigma_{w,\textnormal{LCOGT}}$ &$\mathcal{J}(0.1,5000^2)$ & relative flux (ppm) & Extra jitter term for LCOGT lightcurve. \\
        ~~~$q_{1,\textnormal{LCOGT}}$ &$\mathcal{U}(0,1)$ & --- & Linear limb-darkening coefficient for the LCOGT photometry. \\
        \hline
        RV parameters & & \\
        ~~~$\mu_{\textnormal{CORALIE}}$ &$\mathcal{N}(1220,50^2)$ & m/s & Systemic velocity for CORALIE. \\
        ~~~$\sigma_{w,\textnormal{CORALIE}}$ &$\mathcal{J}(0.1,100)$ & m/s & Extra jitter term for CORALIE. \\
        ~~~$\mu_{\textnormal{HARPS}}$ &$\mathcal{N}(1240,50^2)$ & m/s & Systemic velocity for HARPS. \\
        ~~~$\sigma_{w,\textnormal{HARPS}}$ &$\mathcal{J}(0.1,100)$ & m/s & Extra jitter term for HARPS. \\
        ~~~$\mu_{\textnormal{FIES}}$ &$\mathcal{N}(1215,50^2)$ & m/s & Systemic velocity for FIES. \\
        ~~~$\sigma_{w,\textnormal{FIES}}$ &$\mathcal{J}(0.1,100)$ & m/s & Extra jitter term for FIES. \\       
        \hline
        \hline
    \end{tabular}
    \begin{tablenotes}
      \small
      \item $^1$ To perform the transformation between the $(r_1,r_2)$ plane and the $(b,p)$ plane, we performed the 
      transformations outlined in \cite{Espinoza:2018} depending on the 
      values of $r_{1}$ and $r_{2}$: with $p_l=0$ and $p_u=1$, if $r_{1}>A_r = (p_u-p_l)/(2 + p_l + p_u)$, then $(b,p) = ([1+p_l][1+(r_{1}-1)/(1-A_r)], (1-r_{2})p_l + r_{2}p_u)$. If 
      $r_{1}\leq A_r$, then $(b,p) = ([1+p_l] + \sqrt{r_{1}/A_r}r_{2}(p_u-p_l), p_u + (p_l-p_u)\sqrt{r_{1}/A_r}[1-r_{2}])$.
      \item $^2$ We ensure in each sampling iteration that $e=\mathcal{S}^2_1+\mathcal{S}^2_2 \leq 1$.
      \item $^3$ To transform from the $(q_1,q_2)$ plane to the plane of 
      the quadratic limb-darkening coefficients, $(u_1,u_2)$, we use 
      the transformations outlined in \cite{kipping:2013} for this 
      law $u_1 = 2\sqrt{q_1}q_2$ and $u_2=\sqrt{q_1}(1-2q_2)$.
    \end{tablenotes}
\end{table*}

We now perform the same fits just described, but including the stellar density 
information of \starname\ on our analysis, following our discussion 
in Section \ref{sec:sdmod}. For this, we use the procedures 
described in \cite{Brahm:2018}. Briefly, we first determine the stellar atmospheric
parameters from publicly available HARPS spectra using the \texttt{zaspe} code \citep{zaspe}. We then use a spectral energy distribution model from
\citet{baraffe:2015} having the atmospheric parameters found with \texttt{zaspe},
the public available photometry for K2-140, and its Gaia DR2 parallax \citep{DR2:2018} to determine its stellar radius through an mcmc code\footnote{\url{https://github.com/rabrahm/rstar}}.
Finally, we compute the stellar mass and age by using the Yonsei-Yale stellar
evolutionary models \citep{yi:2001}, which are compared to the spectroscopic effective temperature and the stellar radius via another mcmc code\footnote{\url{https://github.com/rabrahm/isoAR}}.
With this procedure we obtained precise stellar parameters for \starname\ which are presented in  Table \ref{tab:sprops}. 
In particular, we use the estimate of the stellar 
density in the likelihood method 
presented in Section \ref{sec:sdmod} to incorporate this information into 
our modelling. Within \codename, this is also easily done by adding a 
flag to the \codename\ runs. We tried the same fits using $\rho_*$ as a 
parameter instead of $a/R_*$ (a method also dicussed in Section \ref{sec:sdmod}), and obtain the same results as the ones we 
present here.

\begin{table}
    \centering
    \caption{Stellar parameters for \starname.}
    \label{tab:sprops}
    \begin{tabular}{lc} 
        \hline
        \hline
        Atmospheric parameters \\
        ~~~$T_\textnormal{eff}$ (K) & $5736^{+49}_{-45}$ \\[0.1 cm]
        ~~~$\log g$ & $4.479^{+0.016}_{-0.012}$ \\[0.1 cm]
        ~~~$\textnormal{Fe}/\textnormal{H}$ & $0.20 \pm 0.04$ \\[0.1 cm]
        ~~~$v \sin i$ & $2.65 \pm 0.20$ \\[0.1 cm]
        \hline
        Absolute stellar parameters & \\[0.1cm]
        ~~~$M_*$ ($M_\odot$) & $1.077^{+0.020}_{-0.019}$ \\[0.1 cm]
        ~~~$R_*$ ($R_\odot$) & $0.991^{+0.015}_{-0.016}$ \\[0.1 cm]
        ~~~$L_*$ ($L_\odot$) & $0.952^{+0.053}_{-0.052}$ \\[0.1 cm]
        ~~~$\rho_*$ (cgs) & $1.565^{+0.079}_{-0.063}$ \\[0.1 cm]
        ~~~Age (Gyr) & $1.10^{+0.75}_{-1.09}$ \\[0.1 cm]
        ~~~$M_V$ & $4.883^{+0.065}_{-0.071}$ \\[0.1 cm]
        \hline
        \hline
    \end{tabular}
\end{table}

\begin{figure}
   \includegraphics[height=0.75\columnwidth]{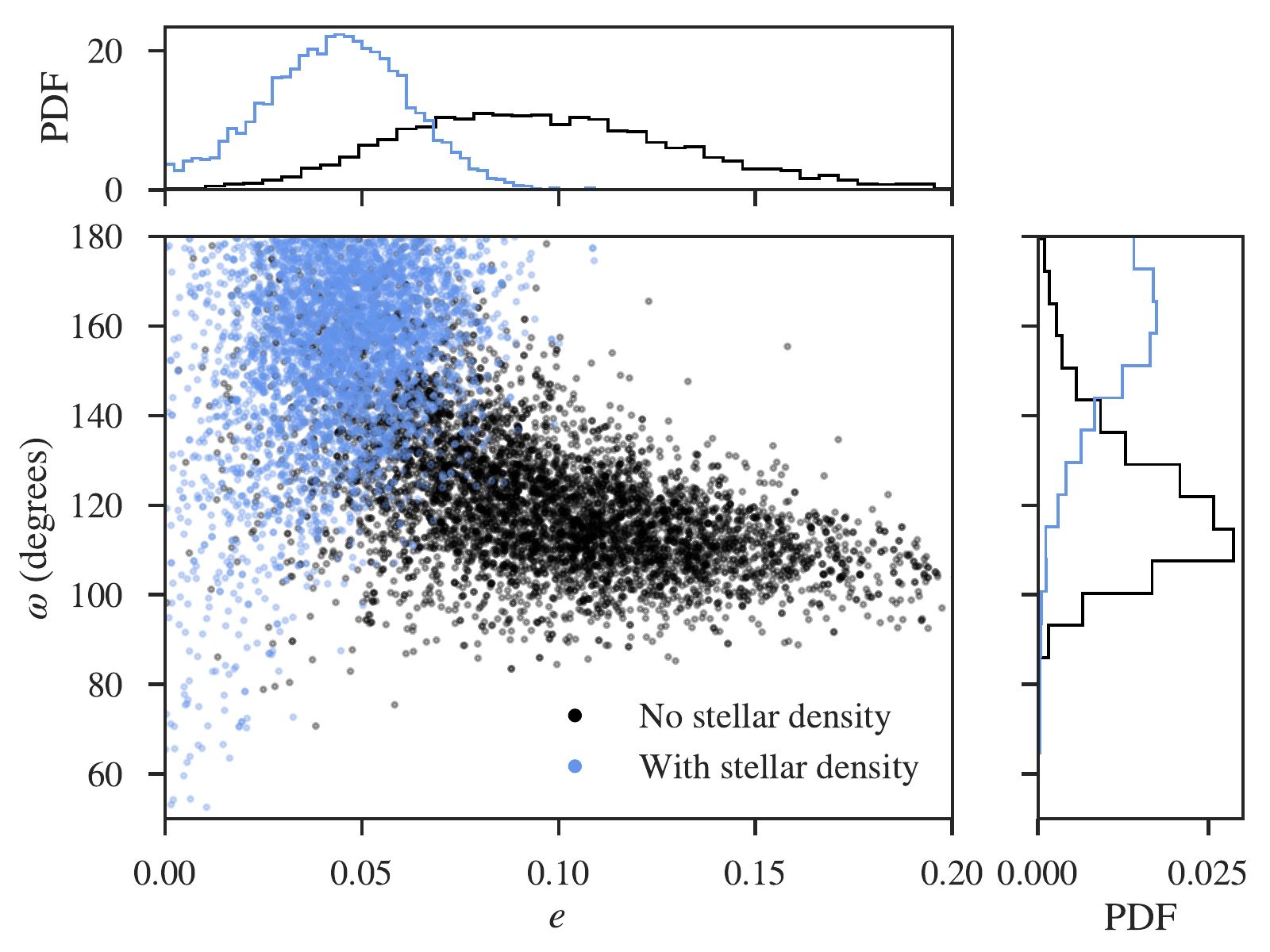}
    \caption{Posterior distribution for the eccentricity and argument of 
    periastron from our joint fits both using (blue) and not (black) the 
    information on the stellar density. Note how this significantly 
    changes the posterior distributions of these quantities.}
    \label{fig:ecc}
\end{figure}

Adding the stellar density information into our analysis significantly 
changes the evidence for eccentricity in the K2-140 system: we 
obtain a $\Delta \ln Z = 1.7$ in favor of a \textit{circular} orbit. 
Although, again, given the current data both models are indistinguishable, 
it is interesting to see how the addition of the stellar density significantly 
changes the evidence now \textit{in favor of a circular orbit}. In fact, 
this addition of the stellar density significantly changes the distribution 
of the eccentricity and argument of periastron between the eccentric joint fit using 
and not using the stellar density information. We show the posterior 
distribution of these parameters for both cases in Figure \ref{fig:ecc}. 
The impact that the stellar density information has on constraining 
the eccentricity and argument of periastron using transits has already 
been made aware by various researchers in the past \citep[see, e.g., ][and references therein]{K:2012,Dawson:2012,K:2014} and has already been 
discussed in length in Section \ref{sec:sdmod}; however, its impact on 
the joint analysis of transits and radial-velocities has only recently 
been started to be exploited thanks to the precise absolute stellar 
parameters that precision spectroscopy combined with the Gaia mission \citep{DR2:2018} provides, and is a property one can take advantage of using \codename. {It is interesting 
to discuss how in Figure \ref{fig:ecc} part of the parameter space in eccentricity and 
the argument of periastron using the stellar density is no longer consistent with the 
values obtained when not using this information. This can be explained as the stellar 
density information gives strong information about the duration of the transit (through 
a strong definition of the parameter $a/R_*$), which in turn constrains the possible 
velocities the planet can acquire based on the transit information. This information 
in turn significantly reduces the parameter space $e$ and $\omega$ can occupy in the 
posterior.}

Our results regarding the eccentric nature of the K2-140b system using all 
the available information (photometry, radial-velocities and the stellar 
density) is, thus, not evident given the data: both models are statistically 
indistinguishable ($\ln Z <2$). However, given the circular model is the 
simpler of the two, we can conclude that this appears to be the best 
model given the data at hand. More data is needed, however, to confidently rule out a significant eccentricity for K2-140b. The 
final transit and radial-velocity parameters obtained with our joint 
fit to the data are presented in Table \ref{tab:k2-140}.

\begin{table}
    \centering
    \caption{Posterior parameters obtained from our joint photometric and radial-velocity \codename\ analysis for \planetnameb\ (including stellar density information).}
    \label{tab:k2-140}
    \begin{tabular}{lc} 
        \hline
        \hline
        Parameter name & Posterior estimate$^a$ \\
                \hline
                \hline
        Posterior parameters for \planetnameb & \\[0.1cm]
        ~~~$P_{b}$ & $6.569298^{+0.000026}_{-0.000027}$ \\[0.1 cm]
        ~~~$t_{0,b}$ (BJD UTC) & $2457588.28381^{+0.00024}_{-0.00024}$ \\[0.1 cm]
        ~~~$a_{b}/R_*$ & $15.24^{+0.13}_{-0.16}$ \\[0.1 cm]
        ~~~$r_{1,b}$ & $0.402^{+0.054}_{-0.047}$ \\[0.1 cm]
        ~~~$r_{2,b}$ & $0.11437^{+0.00083}_{-0.00087}$ \\[0.1 cm]
        ~~~$K_{b}$ (m/s) & $103.8^{+4.8}_{-4.5}$ \\[0.1 cm]
        ~~~$e_b$ & 0 (fixed$^b$, $<0.078$) \\[0.1 cm]
        \hline
        Posterior parameters for K2 photometry & \\[0.1cm]
        ~~~$M_{\textnormal{K2}}$ (ppm) &$-35.5^{+8.4}_{-8.0}$ \\[0.1 cm]
        ~~~$\sigma_{w,\textnormal{K2}}$ (ppm) & $381.7^{+5.7}_{-5.7}$ \\[0.1 cm]
        ~~~$q_{1,\textnormal{K2}}$ &  $0.213^{+0.127}_{-0.075}$ \\[0.1 cm]
        ~~~$q_{2,\textnormal{K2}}$ & $0.57^{+0.22}_{-0.18}$ \\[0.1 cm]
        \hline
        Posterior parameters for LCOGT photometry & \\[0.1cm]
        ~~~$M_{\textnormal{LCOGT}}$ (ppm) &$230^{+230}_{-220}$ \\[0.1 cm]
        ~~~$\sigma_{w,\textnormal{LCOGT}}$ (ppm) & $18.8^{+122.5}_{-16.3}$ \\[0.1 cm]
        ~~~$q_{1,\textnormal{LCOGT}}$ &  $0.586^{+0.064}_{-0.065}$ \\[0.1 cm]
        \hline
        Posterior RV parameters & \\[0.1cm]
        ~~~$\mu_\textnormal{CORALIE}$ (m/s) & ~~~$1214.9^{+7.5}_{-7.9}$ \\[0.1 cm]
        ~~~$\sigma_{w,\textnormal{CORALIE}}$ (m/s) & ~~~$1.16^{+5.3}_{-0.9}$ \\[0.1 cm]
        ~~~$\mu_\textnormal{HARPS}$ (m/s) &  ~~~$1246.6^{+7.3}_{-8.0}$ \\[0.1 cm]
        ~~~$\sigma_{w,\textnormal{HARPS}}$ (m/s) & ~~~$13.5^{+10.3}_{-9.3}$ \\[0.1 cm]
        ~~~$\mu_\textnormal{FIES}$ (m/s) & ~~~~~~$1131.8^{+3.6}_{-3.6}$ \\[0.1 cm]
        ~~~$\sigma_{w,\textnormal{FIES}}$ (m/s) & ~~~$1.6^{+6.7}_{-1.4}$ \\[0.1 cm]
        \hline
        Derived transit parameters for \planetnameb & \\[0.1cm]
        ~~~$R_p/R_*$ & $0.11437^{+0.00083}_{-0.00087}$ \\[0.1 cm]
        ~~~$b = (a/R_*)\cos(i_p)$ & $0.104^{+0.081}_{-0.071}$ \\[0.1 cm]
        ~~~$i_p$ (deg) & $89.60^{+0.27}_{-0.31}$ \\[0.1 cm]
        \hline
        \hline
    \end{tabular}
    \begin{tablenotes}
      \small
      \item $^a$ Errorbars denote the $68\%$ posterior credibility intervals.
      \item $^b$ Upper limits denote the 95\% upper credibility interval of fits when allowing the orbit to be eccentric.
    \end{tablenotes}
\end{table}

\subsection{A \codename\ view of the multi-planetary system around K2-32}
\label{sec:k232}

We now turn our attention to the K2-32 system, a system observed during 
Campaign 2 of the K2 mission in long-cadence mode, which we will model 
using \codename\ in order to showcase various of the features that the code can handle, including the fact that it can efficiently 
fit multi-planetary systems using data from several instruments. 
For this system, we retrieved the radial velocities 
used in \cite{K2-32} which includes radial-velocities obtained in 
that work and in \cite{dai:2016}. For the K2 photometry, we retrieved 
the photometry reduced with the \texttt{EVEREST} pipeline \citep{everest1,everest2} using \texttt{K2DD}\footnote{\url{http://github.com/nespinoza/k2DD}}. The 
\texttt{EVEREST} lightcurves maintain any systematic and astrophysical 
signal in the photometry that is unique to the target, and thus this 
photometry is very useful to showcase 
the ability of \codename\ to model these features (which we here 
model with a GP) along with the transit parameters \textit{and} 
the radial-velocity measurements. As with the K2-140b system, 
the scripts to perform the analyses presented in this section 
are also given in \codename's Github wiki page\footnote{\url{https://github.com/nespinoza/juliet/wiki}}.

K2-32 is a system composed of three exoplanets in a nearly resonant chain 
with periods of 9, 21 and 32 days \citep{K2-32}. In order to model 
it with \codename, we now thus use the stellar density as a fitting 
parameter instead of $a/R_*$ for each planet which, as discussed in 
Section \ref{sec:sdmod}, is the most efficient way of incorporating 
the stellar density information in the fit for multi-planetary 
systems. We used the same method 
as the one for K2-140 described in the previous sub-section to 
estimate the stellar density of K2-32 to be $\rho_* = 2094\pm82$ kg/m$^3$. We fix 
the dilution factor to unity as the dilution from nearby sources based 
on the Gaia detections around 16'' arcseconds from the target ($\sim 4$ K2 pixels)
would all produce dilutions $1>D>0.99$ in $G$, which we assume 
would be similar to the expected dilution in the \textit{Kepler} 
bandpass. 

\begin{figure*}
   \includegraphics[height=1.0\columnwidth]{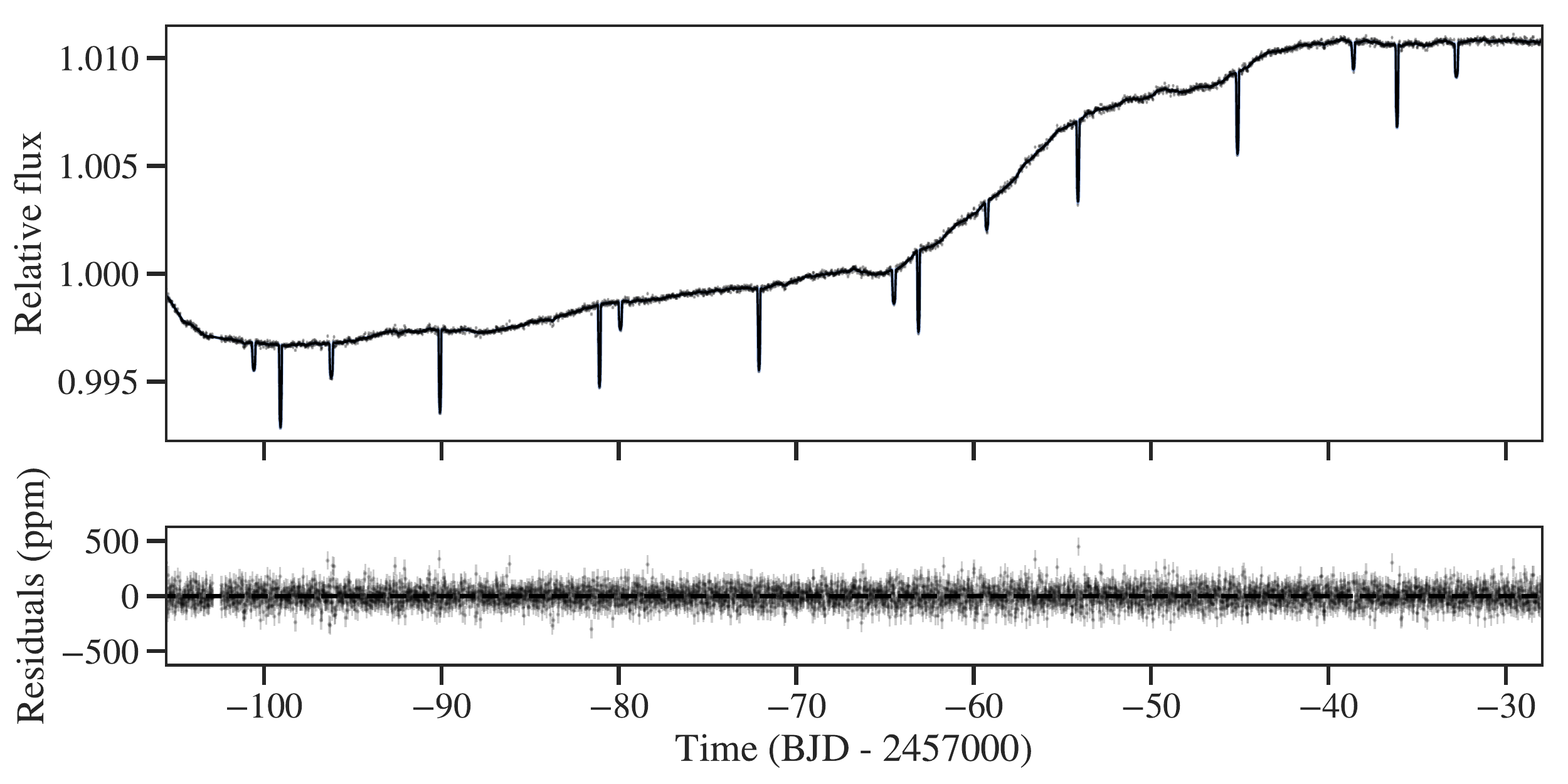}

    \caption{{Top panel.} Full K2 photometry 
    (datapoints with errorbars) for the K2-32 multi-planet system obtained with the \texttt{EVEREST} algorithm \citep{everest1,everest2}. The 
    solid black line indicates the photometric fit component of our 
    \codename\ joint fit to the data, which includes both 
    the transits of the exoplanets K2-32b, K2-32c and K2-32d and the 
    GP used to account for the observed long-term trend. {Bottom panel.} Residuals of our photometric fit.}
    \label{fig:k2-32phot}
\end{figure*}

We tried different GP kernels to account for 
the long-term trend observed in the K2-32 \textit{Kepler} 
photometry: an exponential kernel, the approximate 
Matern kernel introduced in \cite{celerite}, a 
multiplication of those two kernels and the 
quasi-periodic kernel introduced in equation (\ref{eq:celerite}). We performed four model fits 
in total with wide priors for all parameters and 
found via the posterior evidence that the best 
model is the one that uses the multiplication of the 
exponential kernel with the approximate Matern. This 
makes sense with our intuition when looking at the 
light curve (presented in Figure \ref{fig:k2-32phot}): there are not evident quasi-periodic 
oscillations and thus it was unlikely for a quasi-periodic kernel to be a good fit to the data in our case. On the 
other hand, either only an exponential or only a 
Matern kernel would be too strict to account for 
both short term and longer-term trends observed in 
the data. It 
is interesting to note, however, that all 
the fits gave rise to very similar posterior parameters 
for the physical properties of the system (all within 1-sigma), which hints that in this case the selection of the \textit{exact}, 
``best" kernel does not impact too much on the results --- 
it will, however, most likely impact on posterior predictions 
outside of the range of the time-series. We summarize all 
the priors used for the analysis of this system for this 
best-fit scenario in Table \ref{tab:jointk232}. 
For simplicity and ease of comparison, we assume a circular orbit as was assumed 
in \cite{K2-32} --- a full analysis including evidence of eccentric orbits is left 
for future work. In total, we seek the posterior distribution 
of 29 free parameters, and as discussed in the introduction 
of this section, we thus use \texttt{dynesty} to perform 
this fit. As with K2-140b, we also account for the long-cadence integrations 
by supersampling the transit lightcurve with $N=20$ resamples per point with the same 
exposure time used for K2-140b.

\begin{table*}
    \centering
    \caption{Priors used in our joint analysis of the K2-32 multi-planetary system using \codename. The GP used for the K2 photometry was an exponential multiplied by an approximate Matern kernel, both of which were introduced in Section \ref{sec:photmod}.}  
    \label{tab:jointk232}
    \begin{tabular}{lccl} 
        \hline
        \hline
        Parameter name & Prior & Units & Description \\
                \hline
                \hline
        Parameters for K2-32 & & \\
        ~~~$\rho_*$ &$\mathcal{N}(2094,82^2)$ & kg/m$^3$ & Stellar density of K2-32. \\
        \hline
        Parameters for K2-32b & & \\
        ~~~$P_b$ &$\mathcal{N}(6.5693,0.0001^2)$ & days & Period of K2-32b. \\
        ~~~$t_{0,b}$ &$\mathcal{N}(2457588.284,0.001^2)$ & days & Time of transit-center for K2-32b. \\
        ~~~$r_{1,b}$ &$\mathcal{U}(0,1)$ & --- & Parametrization$^{1}$ of \cite{Espinoza:2018} for $p$ and $b$ for K2-32b. \\
        ~~~$r_{2,b}$ &$\mathcal{U}(0,1)$ & --- & Parametrization$^{1}$ of \cite{Espinoza:2018} for $p$ and $b$ for K2-32b. \\
        ~~~$K_{b}$ &$\mathcal{U}(0,100)$ & m/s & Radial-velocity semi-amplitude for K2-32b. \\
        Parameters for K2-32c & & \\
        ~~~$P_c$ &$\mathcal{N}(6.5693,0.0001^2)$ & days & Period of K2-32c. \\
        ~~~$t_{0,c}$ &$\mathcal{N}(2457588.284,0.001^2)$ & days & Time of transit-center for K2-32c. \\
        ~~~$r_{1,c}$ &$\mathcal{U}(0,1)$ & --- & Parametrization$^{1}$ of \cite{Espinoza:2018} for $p$ and $b$ for K2-32c. \\
        ~~~$r_{2,c}$ &$\mathcal{U}(0,1)$ & --- & Parametrization$^{1}$ of \cite{Espinoza:2018} for $p$ and $b$ for K2-32c. \\
        ~~~$K_{c}$ &$\mathcal{U}(0,100)$ & m/s & Radial-velocity semi-amplitude for K2-32c. \\
        Parameters for K2-32d & & \\
        ~~~$P_d$ &$\mathcal{N}(6.5693,0.0001^2)$ & days & Period of K2-32d. \\
        ~~~$t_{0,d}$ &$\mathcal{N}(2457588.284,0.001^2)$ & days & Time of transit-center for K2-32d. \\
        ~~~$r_{1,d}$ &$\mathcal{U}(0,1)$ & --- & Parametrization$^{1}$ of \cite{Espinoza:2018} for $p$ and $b$ for K2-32d. \\
        ~~~$r_{2,d}$ &$\mathcal{U}(0,1)$ & --- & Parametrization$^{1}$ of \cite{Espinoza:2018} for $p$ and $b$ for K2-32d. \\
        ~~~$K_{d}$ &$\mathcal{U}(0,100)$ & m/s & Radial-velocity semi-amplitude for K2-32d. \\
        \hline
        Parameters for K2 photometry & & \\
        ~~~$D_{\textnormal{K2}}$ & 1 (fixed) & --- & Dilution factor for K2. \\
        ~~~$M_{\textnormal{K2}}$ &$\mathcal{N}(0,0.1^2)$ & relative flux & Relative flux offset for K2. \\
        ~~~$\sigma_{w,\textnormal{K2}}$ &$\mathcal{J}(1,1000^2)$ & relative flux (ppm) & Extra jitter term for K2 lightcurve. \\
        ~~~$q_{1,\textnormal{K2}}$ &$\mathcal{U}(0,1)$ & --- & Quadratic limb-darkening parametrization$^{3}$ \citep{kipping:2013}. \\
        ~~~$q_{2,\textnormal{K2}}$ &$\mathcal{U}(0,1)$ & --- & Quadratic limb-darkening parametrization$^{3}$ \citep{kipping:2013}. \\  
        Parameters for the GP of K2 photometry & & \\
        ~~~$\sigma_{\textnormal{K2}}$ &$\mathcal{J}(0.1,10^{4})$ & ppm & Amplitude of the GP. \\
        ~~~$T_{\textnormal{K2}}$ &$\mathcal{J}(0.02,10^5)$ & days & Time-scale of the exponential part of the kernel. \\
        ~~~$\rho_{\textnormal{K2}}$ &$\mathcal{J}(0.02,10^{5})$ & days & Time-scale of the Matern part of the kernel. \\
        \hline
        RV parameters & & \\
        ~~~$\mu_{\textnormal{HIRES}}$ &$\mathcal{N}(0,10^2)$ & m/s & Systemic velocity for HIRES. \\
        ~~~$\sigma_{w,\textnormal{HIRES}}$ &$\mathcal{J}(0.01,10)$ & m/s & Extra jitter term for HIRES. \\
        ~~~$\mu_{\textnormal{HARPS}}$ &$\mathcal{N}(0,10^2)$ & m/s & Systemic velocity for HARPS. \\
        ~~~$\sigma_{w,\textnormal{HARPS}}$ &$\mathcal{J}(0.01,10)$ & m/s & Extra jitter term for HARPS. \\
        ~~~$\mu_{\textnormal{PFS}}$ &$\mathcal{N}(0,10^2)$ & m/s & Systemic velocity for PFS. \\
        ~~~$\sigma_{w,\textnormal{FIES}}$ &$\mathcal{J}(0.01,10)$ & m/s & Extra jitter term for PFS. \\       
        \hline
        \hline
    \end{tabular}
    \begin{tablenotes}
      \small
      \item $^1$ To perform the transformation between the $(r_1,r_2)$ plane and the $(b,p)$ plane, we performed the 
      transformations outlined in \cite{Espinoza:2018} depending on the 
      values of $r_{1}$ and $r_{2}$: with $p_l=0$ and $p_u=1$, if $r_{1}>A_r = (p_u-p_l)/(2 + p_l + p_u)$, then $(b,p) = ([1+p_l][1+(r_{1}-1)/(1-A_r)], (1-r_{2})p_l + r_{2}p_u)$. If 
      $r_{1}\leq A_r$, then $(b,p) = ([1+p_l] + \sqrt{r_{1}/A_r}r_{2}(p_u-p_l), p_u + (p_l-p_u)\sqrt{r_{1}/A_r}[1-r_{2}])$.
      \item $^2$ We ensure in each sampling iteration that $e=\mathcal{S}^2_1+\mathcal{S}^2_2 \leq 1$.
      \item $^3$ To transform from the $(q_1,q_2)$ plane to the plane of 
      the quadratic limb-darkening coefficients, $(u_1,u_2)$, we use 
      the transformations outlined in \cite{kipping:2013} for this 
      law $u_1 = 2\sqrt{q_1}q_2$ and $u_2=\sqrt{q_1}(1-2q_2)$.
    \end{tablenotes}
\end{table*}

\begin{figure}
   \includegraphics[height=2.2\columnwidth]{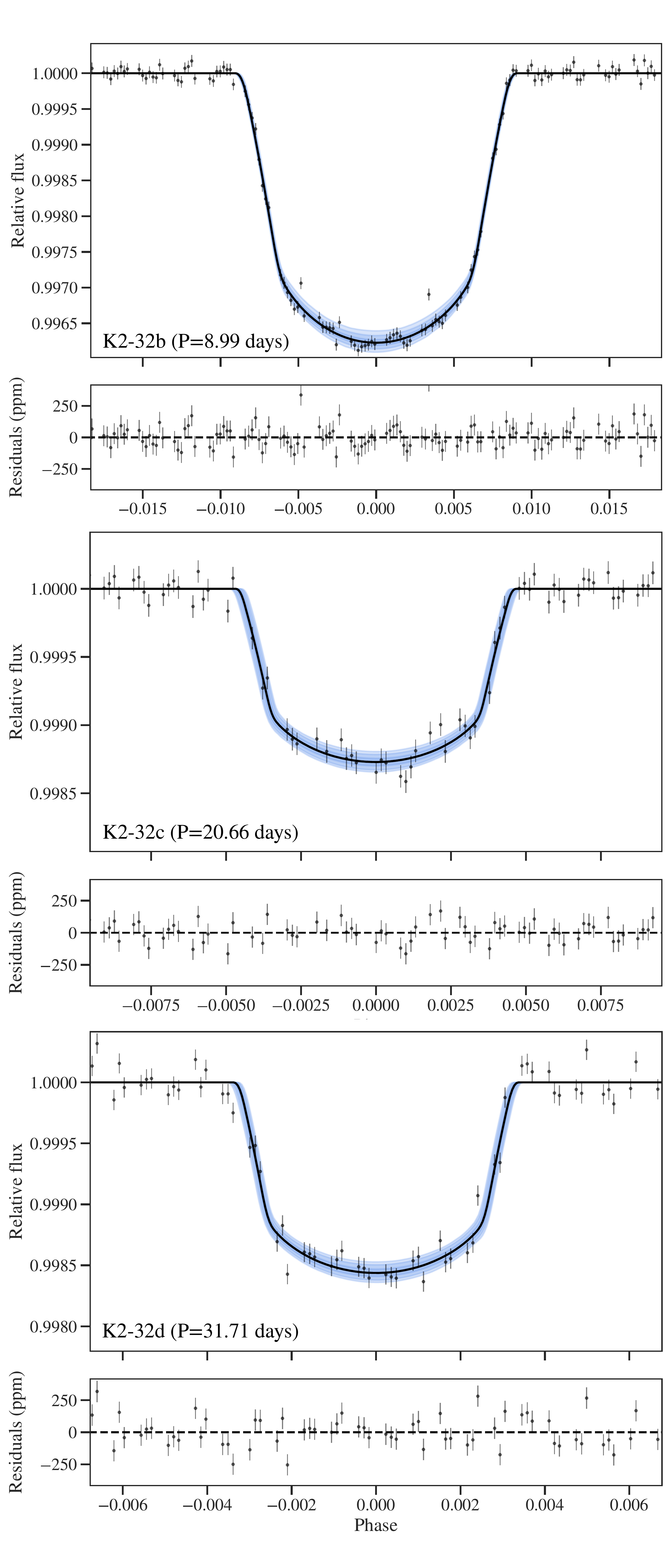}
   \caption{Phased transits after removing the GP component from the 
   K2 photometry for K2-32b (top), K2-32c (middle) and K2-32d (bottom).}
   \label{fig:k2-32phased}
\end{figure}
\begin{figure*}
   \includegraphics[height=1.3\columnwidth]{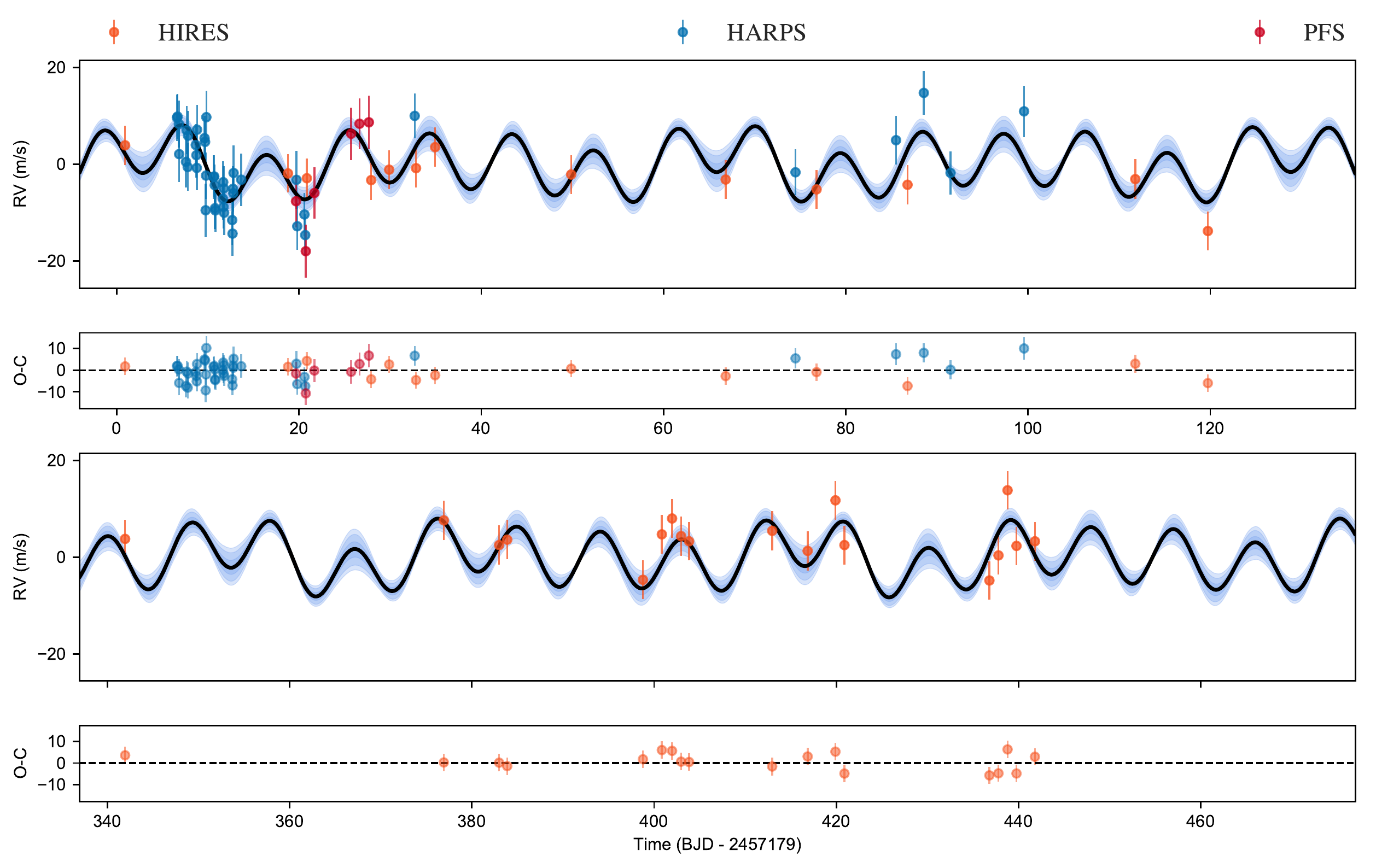}
   \includegraphics[height=0.55\columnwidth]{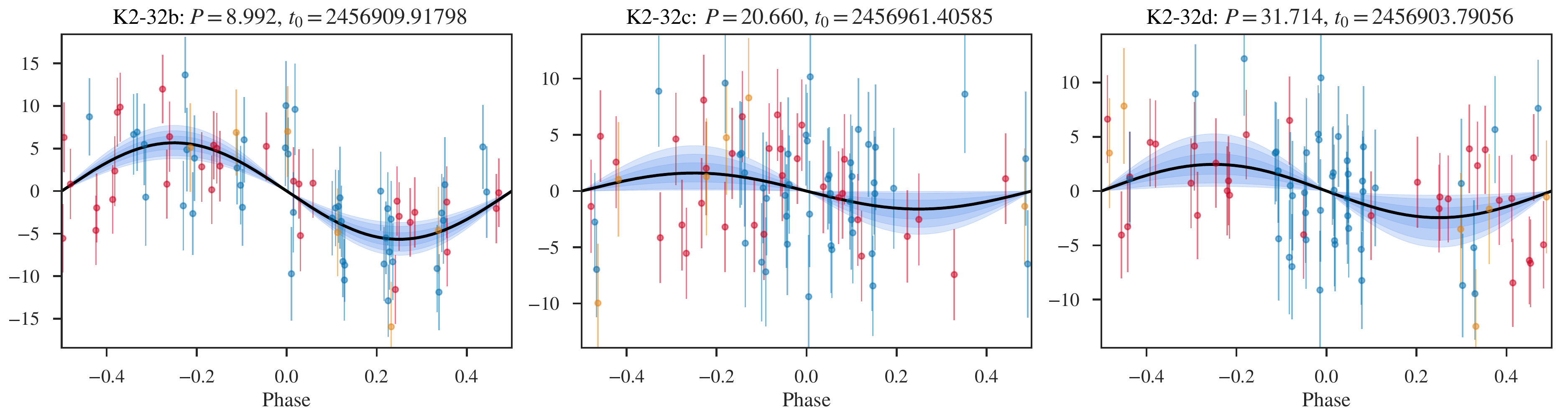}
    \caption{Radial-velocity component of our \codename\ fit to the data 
    of the K2-32 multi-planetary system. The top panel shows the radial-velocities as a function of time and 
    the bottom panels shows the phased radial-velocities after removing 
    the radial-velocity component from the other planets.}
    \label{fig:k2-32rvs}
\end{figure*}

\begin{table}
    \centering
    \caption{Posterior parameters obtained from our joint photometric and radial-velocity \codename\ analysis for the K2-32 multi-planet system.}
    \label{tab:k2-32}
    \begin{tabular}{lc} 
        \hline
        \hline
        Parameter name & Posterior estimate$^a$ \\
                \hline
                \hline
        Posterior parameters for K2-32 & \\[0.1cm]
        ~~~ $\rho_*$ & $2122.6^{+56.9}_{-62.3}$ \\[0.1 cm]
        Posterior parameters for K2-32b & \\[0.1cm]
        ~~~$P_b$ & $8.991854^{+0.000089}_{-0.000093}$ \\[0.1 cm]
        ~~~$t_{0,b}$ (BJD UTC) & $2456909.91797^{+0.00031}_{-0.00028}$ \\[0.1 cm]
        ~~~$r_{1,b}$ & $0.421^{+0.049}_{-0.051}$ \\[0.1 cm]
        ~~~$r_{2,b}$ & $0.05496^{+0.00051}_{-0.00054}$ \\[0.1 cm]
        ~~~$K_{b}$ (m/s) & $5.63^{+0.78}_{-0.76}$ \\[0.1 cm]
        ~~~$e_b$ & 0 (fixed) \\[0.1 cm]
        Posterior parameters for K2-32c & \\[0.1cm]
        ~~~$P_c$ & $20.65974^{+0.00087}_{-0.00085}$ \\[0.1 cm]
        ~~~$t_{0,c}$ (BJD UTC) & $2456961.4058^{+0.0017}_{-0.0016}$ \\[0.1 cm]
        ~~~$r_{1,c}$ & $0.538^{+0.032}_{-0.037}$ \\[0.1 cm]
        ~~~$r_{2,c}$ & $0.03108^{+0.00054}_{-0.00056}$ \\[0.1 cm]
        ~~~$K_{c}$ (m/s) & $1.68^{+0.83}_{-0.82}$ \\[0.1 cm]
        ~~~$e_c$ & 0 (fixed) \\[0.1 cm]
        Posterior parameters for K2-32d & \\[0.1cm]
        ~~~$P_d$ & $31.7143^{+0.0013}_{-0.0011}$ \\[0.1 cm]
        ~~~$t_{0,d}$ (BJD UTC) & $2456903.7905^{+0.0016}_{-0.0018}$ \\[0.1 cm]
        ~~~$r_{1,d}$ & $0.63156^{+0.02001}_{-0.02124}$ \\[0.1 cm]
        ~~~$r_{2,d}$ & $0.03673^{+0.00059}_{-0.00064}$ \\[0.1 cm]
        ~~~$K_{d}$ (m/s) & $2.43^{+0.94}_{-0.87}$ \\[0.1 cm]
        ~~~$e_d$ & 0 (fixed) \\[0.1 cm]
        \hline
        Posterior parameters for K2 photometry & \\[0.1cm]
        ~~~$M_{\textnormal{K2}}$ (ppm) &$-3500^{+860}_{-870}$ \\[0.1 cm]
        ~~~$\sigma_{w,\textnormal{K2}}$ (ppm) & $83.6^{+1.1}_{-1.1}$ \\[0.1 cm]
        ~~~$q_{1,\textnormal{K2}}$ & $0.334^{+0.155}_{-0.104}$ \\[0.1 cm]
        ~~~$q_{2,\textnormal{K2}}$ & $0.61^{+0.19}_{-0.16}$ \\[0.1 cm]
        Posterior GP parameters for K2 photometry & \\[0.1cm]
        ~~~$\sigma_{\mathcal{GP},\textnormal{K2}}$ (ppm) & $10.46^{+7.3}_{-4.3}$ \\[0.1 cm]
        ~~~$T_\textnormal{K2}$ (days) & $16647^{+30333}_{-10917}$ \\[0.1 cm]
        ~~~$\rho_\textnormal{K2}$ (days) & $49^{+24}_{-16}$ \\[0.1 cm]
        \hline
        \hline
        Posterior RV parameters & \\[0.1cm]
        ~~~$\mu_\textnormal{HIRES}$ (m/s) & ~~~ $-1.722^{+0.714}_{-0.705}$ \\[0.1 cm]
        ~~~$\sigma_{w,\textnormal{HIRES}}$ (m/s) & ~~~ $3.65^{+0.65}_{-0.56}$ \\[0.1 cm]
        ~~~$\mu_\textnormal{HARPS}$ (m/s) &  ~~~ $1.092^{+0.701}_{-0.706}$ \\[0.1 cm]
        ~~~$\sigma_{w,\textnormal{HARPS}}$ (m/s) & ~~~ $3.92^{+0.67}_{-0.58}$ \\[0.1 cm]
        ~~~$\mu_\textnormal{PFS}$ (m/s) & ~~~ $-6.5^{+2.0}_{-2.0}$ \\[0.1 cm]
        ~~~$\sigma_{w,\textnormal{PFS}}$ (m/s) & ~~~ $4.6^{+2.2}_{-1.8}$ \\[0.1 cm]
        \hline
        \hline
    \end{tabular}
    \begin{tablenotes}
      \small
      \item $^a$ Errorbars denote the $68\%$ posterior credibility intervals.
    \end{tablenotes}
\end{table}

\begin{table}
    \centering
    \caption{Derived parameters from our joint photometric and radial-velocity \codename\ analysis for the K2-32 multi-planet system.}
    \label{tab:k2-32-2}
    \begin{tabular}{lc} 
        \hline
        \hline
        Parameter name & Posterior estimate$^a$ \\
                \hline
                \hline
        Derived parameters for K2-32b & \\[0.1cm]
        ~~~$R_{p,b}/R_*$ & $0.05497^{+0.00052}_{-0.00054}$ \\[0.1 cm]
        ~~~$b_b = (a_b/R_*)\cos(i_{p,b})$ & $0.135^{+0.074}_{-0.076}$ \\[0.1 cm]
        ~~~$(a_b/R_*)$ & $20.85^{+0.18}_{-0.20}$ \\[0.1 cm]
        ~~~$i_{p,b}$ (deg) & $89.62^{+0.21}_{-0.21}$ \\[0.1 cm]
        Derived parameters for K2-32c & \\[0.1cm]
        ~~~$R_{p,c}/R_*$ & $0.03108^{+0.00054}_{-0.00056}$ \\[0.1 cm]
        ~~~$b_c = (a_c/R_*)\cos(i_{p,c})$ & $0.307^{+0.048}_{-0.056}$ \\[0.1 cm]
        ~~~$(a_c/R_*)$ & $36.31^{+0.32}_{-0.35}$ \\[0.1 cm]
        ~~~$i_{p,c}$ (deg) & $89.51571^{+0.08934}_{-0.08008}$ \\[0.1 cm]
        Derived parameters for K2-32d & \\[0.1cm]
        ~~~$R_{p,d}/R_*$ & $0.03673^{+0.00059}_{-0.00064}$ \\[0.1 cm]
        ~~~$b_d = (a_d/R_*)\cos(i_{p,d})$ & $0.44734^{+0.03001}_{-0.03186}$ \\[0.1 cm]
        ~~~$(a_d/R_*)$ & $48.32^{+0.42}_{-0.47}$ \\[0.1 cm]
        ~~~$i_{p,d}$ (deg) & $89.470^{+0.038}_{-0.038}$ \\[0.1 cm]
        \hline
        \hline
    \end{tabular}
    \begin{tablenotes}
      \small
      \item $^a$ Errorbars denote the $68\%$ posterior credibility intervals.
    \end{tablenotes}
\end{table}

Figures \ref{fig:k2-32phot}, \ref{fig:k2-32phased} and \ref{fig:k2-32rvs} 
show the results of our joint fit to the data, which reveals an 
excellent fit to the whole dataset. As can be observed in Figure 
\ref{fig:k2-32phot}, the GP we used to account for the long-term 
trend captures perfectly well the observed long-term trend in the 
K2 photometry. We show the phased transits of the exoplanets K2-32b, K2-32c and K2-32d after substracting this GP component 
from the data in Figure \ref{fig:k2-32phased}. In 
Figure \ref{fig:k2-32rvs} we present the radial-velocity 
component of our joint fit, which shows a very similar shape 
as the radial-velocity only analysis presented in \cite{K2-32}. 
As can be seen, thus, \codename\ can efficiently fit 
multi-planetary systems.

We present the resulting posterior parameters of the system with 
our \codename\ joint fit in Tables \ref{tab:k2-32} and \ref{tab:k2-32-2}. Comparing our posterior parameters with the 
ones published by \cite{K2-32}, we see that we obtain values 
in excellent agreement albeit more precise than that previous 
work. This is most likely a result of the fact that our joint analysis including the stellar density provides more precise 
timing ephemerides, which in turn lets us obtain more 
precise values for the semi-amplitudes of the planets in 
the system.

\section{Discussion}
\label{sec:dc}

As was presented in Section \ref{sec:tests}, \codename\ is a very flexible code 
that allows to incorporate a variety of setups in the analysis of photometry, 
radial-velocity or both, for both transiting and non-transiting 
systems, as was illustrated with our analysis of the K2-140 
system in Section \ref{sec:k2140} and the multi-planetary system 
around K2-32b in Section \ref{sec:k232}. It is efficient both 
at exploring wide parameter spaces, and also at providing 
quantitative measures of evidence of adding or not 
extra parameters/models on the fits (e.g., dilution factors, 
GPs, eccentricity, additional planets in the system). In 
this Section, we discuss and explore how the features 
provided by \codename\ can be used in the analysis of other 
datasets, along with a discussion on the speed of the {library} 
in different settings.

\subsection{GP hyperparameter sharing within \codename}
In Sections \ref{sec:k2140} and \ref{sec:k232}, we showed how 
GPs can be introduced to different instruments independently 
in the photometry. However, one feature we did not present 
here but which is also available within \codename\ is that, 
as explained in Sections \ref{sec:photmod} and \ref{sec:rvmod}, 
the hyperparameters of these kernels can be shared not only with 
a GP being incorporated in the radial-velocity dataset, but they 
can also be shared \textit{between} photometric datasets as well. {This feature of \codename\ has been used 
in \cite{luque:2019}, for example, to estimate the rotation 
period of the star GJ 357 using photometry from different 
ground-based instruments which were fit with a quasi-periodic 
kernel, where the characteristic period and time-scale of the 
process were common to all instruments, but the amplitudes and 
jitter terms were not (due to, e.g., different passbands). 
This allowed to retrieve a very precise estimate of this 
parameter which was critical for the study of the stellar properties and their relation to activity.} If in addition 
to these datasets a radial-velocity dataset is also used where signatures of rotational 
modulation are either suspected or observed, a GP can also be fit simultaneously to 
the radial-velocities which, as already mentioned, can in 
turn also share said hyperparameters of the 
photometric GPs. We 
believe this kind of joint analyses are not only important for transiting exoplanets, 
but could also be key for radial-velocity analyses in order to correctly propagate 
the information between the photometry and the radial-velocities in a consistent 
way, especially in cases where the rotation period of stars (usually estimated 
through photometric rotational modulation) are very close, 
fractions, or at the periods of 
suspected planets in radial-velocity analyses \citep[see, e.g.,][]{tuomi,diaz}.

\subsection{\codename\ as a planet detection tool}
As it was briefly introduced in Section \ref{sec:rvtest}, 
\codename\ can also be used as a planet detection tool 
similar to \texttt{kima}, a tool presented by \cite{kima:2018} 
for the detection of exoplanets in radial-velocity datasets. 
\texttt{kima}, however, is more efficient than \texttt{juliet} 
at this task as it includes the number of planets in the system 
as a free parameter itself, performing thus one fit instead of 
the many fits for different models that one has to perform 
with \texttt{juliet}. {However, the versatility 
\texttt{juliet} offers in terms of kernel types for modelling 
stellar activity might make it a good competitor. Indeed, 
\codename\ has already been used in \cite{espinoza:2019}, \cite{kos:2019}, \cite{brahm:2019} and \cite{luque:2019} to 
search for additional planets in radial-velocities, tightly constraining the presence of 
at least three planets in the latter work. \codename\ } might 
{also} be seen as more versatile in the sense that it 
can not only handle different instruments but also transits, 
if available, which in turn can help not only constrain a 
subset of the suspected exoplanets in the system embedded in 
a given radial-velocity dataset if their transits are observed 
in the photometry, but also help in the search for evidence of 
transiting exoplanets in photometric data alone. This latter 
usage of \codename\ has already been introduced in \cite{espinoza:2019} when searching for 
transits of TOI-141c, and we believe it might be interesting 
to compare against simpler (but faster) algorthms such as the 
widely used Box-Least Squares \citep[BLS;][]{bls:2002} 
algorithm. In particular, it will be interesting to test 
if the error underestimation of the bayesian evidences 
discussed in \cite{Nelson:2018} when quantifying the evidence 
of additional planets in radial-velocity datasets is also a 
problem on the search for additional planets in transit 
photometry.

\subsection{Computing speed of \codename}
In this work we have made many analyses with \codename\ 
but we have not discussed yet how much it takes for the fits 
to converge (see Section \ref{sec:sampling}), which might be 
one of the key points that might define whether one wants to 
use \codename\ or other open source tools like the ones 
discussed in the introduction of this manuscript for the 
problem at hand. In general, the smaller the prior, the 
faster the algorithm will converge {for obvious 
reasons already discussed in Section \ref{sec:sampling}}. 
On this front, it is important to note 
that in this work we deliberately tried very 
wide priors for all the parameters to showcase the 
ability of nested samplers to properly explore the whole parameter space. Most of the analyses made 
in this work were performed using a laptop with an Intel(R) Core(TM) 
i5-7287U CPU at 3.30GHz. In this laptop, for single 
photometry and/or radial-velocity analyses \codename\ takes of order 
of minutes depending on the complexity of the fit. For multiple-instrument 
photometry or radial-velocity fits without the inclusion of GPs, \codename\ 
takes on the order of minutes to converge with the standard 
\texttt{MultiNest} and \texttt{dynesty} options built in 
within \codename. When GPs are included, the runs tested 
here took on the order of tens of minutes to converge. 

For joint analyses of photometry and radial-velocities, 
for a single planetary system with only one instrument and no GPs, 
\codename\ takes tens of minutes to run. However, as the dimensions 
and complexity of the problems increase, the computing speed increases 
as well. For example, the K2-140b full analysis presented in Section \ref{sec:k2140} took several hours to converge in the laptop mentioned 
above, and we found, as expected, that the speed of convergence strongly 
depends on the priors (larger priors take longer to converge). As a rule 
of thumb, problems with dimensions of order $\sim 20$ take several 
hours (depending on the complexity of the problem, e.g., if several GPs 
are included in the fit it could take of order a day in a laptop like 
the one defined above). However, when approaching these large number 
of dimensions, users might want to use the multi-threading 
capabilities \codename\ provides through \texttt{MultiNest} and 
\texttt{dynesty}. For example, the full fit of the K2-32 system detailed 
in Section \ref{sec:k232}, which includes 29 free parameters and a GP in 
time to model the K2 systematics took only 1 hour when \codename\ was 
ran in multi-threading mode using \texttt{dynesty} with 10 cores on a 
Intel(R) Xeon(R) CPU E5-2699 v3 at 2.30GH machine. Within 
\codename\, multi-threading for \texttt{dynesty} can be included with 
a simple flag that determines the number of threads one wants to use. 
Via \texttt{MultiNest}, direct support of OpenMPI is available as 
\texttt{PyMultiNest} automatically recognizes this call.

\section{Conclusions and Future Work}
\label{sec:cfw} 
In this work we have presented a new versatile open source code, 
\codename, with which one can perform efficient fitting of 
photometry, radial-velocity or both, allowing the user to not 
only thoroughly explore the parameter space but which also 
provides estimates for the bayesian model evidence (thanks to 
nested sampling algorithms), with which one can formally perform 
model comparison. Using two previously analyzed systems, K2-140b \citep{Giles:2018,Korth:2019} and K2-32b \citep[see][and references 
therein]{K2-32}, we have shown how \codename\ is versatile enough to 
allow for multiple-instruments, dilutions, GPs and even 
multi-planetary systems to be efficiently fit, allowing in turn to 
answer questions in terms of, e.g., the evidence of eccentricity in a given dataset, the best kernel to use 
for a GP in a given dataset or even the presence of 
additional signals both in transits and in radial-velocities. As such, \codename\ is a 
versatile tool for the characterization of extrasolar planets 
that we believe will prove to be very useful in the \textit{TESS} 
era of planet discovery.

Our plans for future work are multiple. In the near-future, we plan 
to incorporate support for secondary eclipses which can easily be 
implemented within \texttt{batman} \citep{batman:2015}, and thus 
allow users to optionally include this effect in the joint 
and/or photometric fitting with \codename. We also plan to 
incorporate within \codename\ other codes that allow to model 
a plethora of other photometric effects such as \texttt{starry} \citep{starry} 
and/or \texttt{spiderman} \citep{spiderman}. This would 
in turn transform \codename\ into a characterization 
toolbox which will not only be able to discover the 
exoplanets under study, but also shine some light on 
what these exoplanets are made of. 

On the front of GP regression, we aim at implementing GP kernels on 
request. We have made sure to incorporate in this first version of 
\codename\ the most popular GP kernels in use but it might happen 
that other kernels gain popularity in the near future. Implementing 
new kernels within \codename\ is relatively easy given the 
excellent work that has been made on the packages used by the 
code to implement GPs (\texttt{george} and \texttt{celerite}) and 
as such we believe that not only us but the community could 
perform their own kernel implementations and push them to the 
\codename\ Github repository. 

We note that a full documentation for \codename\ has been 
published alongside this paper\footnote{\url{https://juliet.readthedocs.io/en/latest/}}.

\section*{Acknowledgements}
{We would like to thank the referee, Daniel Foreman-Mackey, 
for very useful suggestions that greatly improved the paper 
and the \codename\ library}. N.E. would like to thank the Gruber Foundation for its generous support to this research. N.E. would like to thank J. Buchner and J. Speagle for 
useful discussions regarding the usage of MultiNest and 
\texttt{dynesty}, respectively. N.E. and D.K. would like to thank 
M. Gunther and T. Daylan for fruitful discussions on the topics 
discussed in this article and T. Trifonov {and the CARMENES 
team on discussions regarding} the analysis of radial-velocities. R.B.\ acknowledges 
support from FONDECYT Post-doctoral Fellowship Project 3180246, and from the Millennium Institute of Astrophysics (MAS).





\bibliographystyle{mnras}
\bibliography{paperbib}


\bsp	
\label{lastpage}
\end{document}